\documentclass{aa}

\usepackage{graphicx}
\usepackage{txfonts}
\usepackage{hyperref}
\usepackage{amsmath}
\usepackage{ulem}

\begin{document}

\title{Assembly bias from nuisance to probe}
\subtitle{II: Salvaging linear clustering from DESI and SDSS data}
\titlerunning{Assembly bias from nuisance to probe II}

\author{
Nelson Padilla\inst{\ref{Iate},\ref{OAC}}\and
Dante Paz\inst{\ref{Iate},\ref{OAC}} \and
Ivan Lacerna\inst{\ref{uda}}
}  

\institute{
CONICET. Instituto de Astronomía Teórica y Experimental (IATE). Laprida 854, Córdoba X5000BGR, Argentina \label{Iate}\\
\email{nelson.padilla@unc.edu.ar}
\and 
Universidad Nacional de Córdoba (UNC). Observatorio Astronómico de Córdoba (OAC). Laprida 854, Córdoba X5000BGR, Argentina\label{OAC}
\and
Instituto de Astronom\'ia y Ciencias Planetarias, Universidad de Atacama, Copayapu 485, Copiap\'o, Chile \label{uda} 
}

\date{\today}

\abstract
{Galactic conformity links the properties of neighbouring galaxies and is
usually interpreted as a signature of assembly-biased galaxy occupation. We
study projected and redshift-space compensated conformity statistics in
colour-selected DESI BGS and SDSS MGS samples. These statistics combine
correlations of related galaxy populations so that shared nonlinear clustering
contributions are partially suppressed while a differential large-scale
response can remain.}
{We test whether this suppression exposes the shape of the linear matter
correlation function on scales where ordinary colour-selected clustering has a
scale-dependent nonlinear response.}
{We split galaxies into red and blue subsamples and measure ordinary
correlations, conformity statistics, and compensated combinations in projection
and in redshift-space monopoles. Our default measurement uses central
primaries. We fit three-dimensional and projected linear matter correlation
function templates, \(\xi_{\rm mm}^{\rm lin}\) and
\(w_{\rm mm}^{\rm lin}\), diagnose the response with effective kernels, and
compare with MTNG, FLAMINGO, and MDPL2--SAG models.}
{Ordinary high-colour clustering broadly follows the linear matter templates
but develops a strongly scale-dependent response at small separations. In
DESI, the conformity monopole \(\Delta f_0(s)\), projected
\(\Delta f(r_p)\), and compensated combinations such as \(C_w(r_p)\)
track the linear matter shape substantially further into the nonlinear regime,
consistent with suppression of nonlinear clustering modes while
leaving a linear-matter-like mode visible. The effect is stronger for
central-primary and dense samples. Projected effective kernels are enhanced at
low line-of-sight separations, helping the projected residual resemble
\(\xi_{\rm mm}^{\rm lin}(r_p)\) better than
\(w_{\rm mm}^{\rm lin}(r_p)\). SDSS is consistent but noisier, while the
simulations reproduce the qualitative behaviour with model-dependent
amplitudes and residual scale dependence.}
{{Compensated conformity statistics provide small-scale observables that follow the linear matter correlation-function shape despite the presence of nonlinear clustering. This behaviour is consistent with the suppression of scale-dependent nonlinear contributions.} In DESI BGS this behaviour is
strongest for dense low-redshift samples and central-primary definitions.
The results suggest that assembly-sensitive population differences can filter
nonlinear clustering modes, although the surviving nonlinear contamination must
be calibrated before precision cosmological use.}


\maketitle

\section{Introduction}
\label{sec:introduction}

Galaxy properties depend on environment.  This is reflected in the
morphology--density relation, in environmental trends of colour and
star-formation activity, and in marked clustering measurements that connect
galaxy populations to the underlying large-scale structure
\citep{Dressler1980Morphology,Kauffmann2004Env,sheth2005marked,
skibba2006marked}.  A particularly direct expression of this connection is
galactic conformity: galaxies around quenched or passive primaries are
themselves more likely to be quenched than galaxies around star-forming
primaries \citep{Weinmann2006,Kauffmann2013Conformity}.  On scales beyond the
virial radius, two-halo conformity has often been interpreted as a manifestation
of assembly bias, namely the dependence of halo clustering on secondary halo
properties at fixed mass
\citep{Wechsler2006,GaoWhite2007,Hearin2015Beyond,calderon2018conformity,
pahwa2017conformity,LacernaPadillaPalma2025}.  In this picture, galaxy
properties respond to halo assembly, while halo assembly retains memory of the
larger-scale density field.  The same effect that complicates the use of
galaxies as matter tracers can therefore become a probe of the connection
between galaxy formation and large-scale structure.

The first paper in this series \citep{PadillaLacernaPaz2026Letter}
used TNG300 galaxies in real space to show that
the usual conformity statistic
\begin{equation}
  \Delta f_{\rm Q}(r)
  \equiv
  f_{\rm Q}(r|{\rm Q}) -
  f_{\rm Q}(r|{\rm SF})
\end{equation}
can be written in terms of auto- and cross-correlations of
quenched and star-forming samples.
On sufficiently large scales, differences in the clustering
biases of the two populations produce a conformity signal
whose scale dependence follows the linear matter correlation
function, $\xi_{\rm mm}^{\rm lin}(r)$.
{We use the term effective clustering response to describe
the amplitude and scale dependence of a clustering statistic
relative to $\xi_{\rm mm}^{\rm lin}$. A constant response
indicates that the statistic follows the linear matter
correlation function up to an overall amplitude, whereas
a scale-dependent response indicates departures from this
shape.}
Paper I found that the agreement with the linear template
extended to smaller scales as the galaxy number density
increased, and that the amplitude of the signal could not
be explained by halo-mass bias alone
\citep{PadillaLacernaPaz2026Letter}.
This motivates the present analysis: dense low-redshift
samples may allow conformity-like combinations of galaxy
correlations to preserve the linear matter correlation
function shape while suppressing part of the nonlinear
galaxy-clustering contribution.

We focus on binary splits of the galaxy population and construct
compensated combinations of their auto- and cross-correlations.
{The two populations trace the same underlying matter
distribution, but generally with different clustering biases
and nonlinear contributions. Combining their correlations
with opposite signs or through ratios can suppress contributions
that enter with similar relative amplitudes.
Importantly, this cancellation can affect both linear and
nonlinear clustering. Differences in the large-scale biases
of the populations can nevertheless leave a residual signal
proportional to $\xi_{\rm mm}^{\rm lin}$. If nonlinear
contributions cancel more efficiently than this linear
signal, the effective clustering response can remain
approximately constant on scales where ordinary galaxy
clustering becomes nonlinear.}
We test whether this behaviour occurs in observed and
simulated galaxy populations.

This approach connects two long-standing uses of low-redshift galaxy
clustering.  Measurements in the {2-degree Field Galaxy Redshift Survey (2dFGRS) and Sloan Digital Sky Survey (SDSS)} established that clustering
depends strongly on luminosity, colour, spectral type, and stellar population
\citep{Norberg2001,Norberg2002,Hawkins2003,Zehavi2005,Zehavi2011,
Guo2013CMASS}.  At the same time, broad-band galaxy clustering has been used
to constrain the matter-density scale, the transfer-function shape, and the
combination of galaxy bias and matter clustering amplitude
\citep{Peacock2001,Percival2001,Tegmark2004PowerSpectrum,Tegmark2004,
Cole2005,Sanchez2006,Tegmark2006LRG}.  The tension between these facts is the
opportunity exploited here: galaxy bias complicates cosmological
interpretation, but differences between galaxy populations
also make it possible to construct combinations in which
some clustering contributions cancel while a differential
large-scale signal remains.

{Our approach shares with multi-tracer methods
\citep{McDonaldSeljak2009,Hamaus2010} the use of different
galaxy populations to construct combinations that suppress
shared contributions. However, multi-tracer methods exploit
differences in tracer bias to reduce cosmic variance, whereas
our combinations are designed to reduce departures from
linear matter clustering. We do not assume cosmic-variance
cancellation.} Related ideas involving combinations of
tracers also appear in bias expansions
\citep{DesjacquesJeongSchmidt2018} and
clustering--lensing statistics
\citep{Baldauf2010,Mandelbaum2013}.

The main observational sample is the {Dark Energy Spectroscopic Instrument (DESI)} Bright Galaxy Survey (BGS), complemented by the SDSS Main Galaxy Sample (MGS); together, they provide the high-number-density samples advocated in \citet{PadillaLacernaPaz2026Letter}.
Dense low-redshift samples are particularly useful because the signal is built
from differences between neighbour fractions and cross-correlations and
therefore benefits from high number density
\citep{Strauss2002MGS,Zehavi2011,RuizMacias2020BGS,Ross2024DESILSS,
DESIDR2BAO2025}.

We compare the observations with MillenniumTNG, FLAMINGO, and MDPL2--SAG
\citep{Pakmor2023MTNGHydro,Bose2023MTNGClustering,Schaye2023FLAMINGO,
Kugel2023FLAMINGOCalibration,Cora2018SAG,Knebe2018MultiDarkGalaxies}.  These
catalogues extend the real-space TNG300 test of Paper I to larger volumes,
redshift space, and projected measurements.  We use them as comparative
testbeds, not as calibrated models of DESI or SDSS colour-dependent clustering.

{We also explore broad-band shape sensitivity by varying
the effective matter-density parameter
$\Omega_m^{\rm shape}$, which controls the shape of
the linear matter correlation-function template
within a fixed cosmological model, including the
corresponding Alcock--Paczynski (AP) remapping.
The procedure is described in
Appendix~\ref{app:omega_shape}.}

The paper is organized as follows.  Sect.~\ref{sec:theory} defines the
statistics.  Sects.~\ref{sec:data} and \ref{sec:sims} describe the observed and
simulated samples.  Sect.~\ref{sec:results} presents ordinary clustering,
conformity statistics, compensated statistics, and effective kernels.  We
discuss the implications in Sect.~\ref{sec:conclusions}.  Appendices
\ref{app:nulls} and \ref{app:omega_shape} summarize auxiliary compensated
combinations and exploratory effective-shape fits.

\section{Compensated conformity statistics}
\label{sec:theory}

We use ``compensated'' by analogy with compensated filters, but here the
cancellation operates between galaxy samples instead of spatial scales.
Correlations of populations with similar scale dependence enter with opposite
signs, leaving a residual set by differences in their clustering response.

{We begin with a parent galaxy sample defined by the survey
selection and number-density cut, which we divide at the
median colour into disjoint high- and low-colour populations.
We distinguish primary galaxies, around which
neighbour statistics are measured, from neighbours, which
are drawn from the full parent sample and include
both centrals and satellites. By default, primaries are
restricted to galaxies identified as centrals by the group
finder (central-primary selection). We also consider
all-primary selection, in which both centrals and satellites
can serve as primaries. The neighbour sample is identical
in both configurations. Central-primary selection removes satellites from
the primary sample but does not eliminate one-halo pairs,
since satellites remain in the neighbour sample.}

\subsection{Binary samples and projected conformity}
\label{sec:binary_samples}

We split the {full neighbour sample, denoted by ${\rm g}$,} into two disjoint
populations, $A$ and $B$, with number densities
$\bar n_A$ and $\bar n_B$. We define
\begin{equation}
  \bar n_{\rm g}=\bar n_A+\bar n_B,
  \qquad
  \bar f_A=\frac{\bar n_A}{\bar n_{\rm g}} .
\end{equation}
The corresponding primary populations are denoted
$H$ and $L$, with central-primary selection used
by default.

For a primary sample \(P\in\{H,L\}\) and a neighbour sample
\(Y\in\{A,B\}\), we define
\begin{equation}
  w_{PY}(r_p)
  =
  2\int_0^{\pi_{\rm max}}
  \xi_{PY}(r_p,\pi)\,d\pi ,
  \qquad
  \omega_{PY}(r_p)
  =
  \frac{w_{PY}(r_p)}{2\pi_{\rm max}} ,
  \label{eq:wpy_def}
\end{equation}
where {$r_p$ and $\pi$ are the separations perpendicular and
parallel to the line of sight, respectively.}
The finite value of \(\pi_{\rm max}\) is part of the observable definition, and
\(1+\omega_{PY}\) is the projected pair-count factor entering neighbour
fractions. {We adopt \(\pi_{\rm max}=80~h^{-1}\) Mpc throughout.}

The projected fraction of \(A\)-type neighbours around primaries of type \(P\)
is
\begin{equation}
  f_A(r_p|P)
  =
  \bar f_A
  \frac{1+\omega_{PA}(r_p)}
       {1+\omega_{P{\rm g}}(r_p)}
  =
  \frac{
    \bar n_A[1+\omega_{PA}(r_p)]
  }{
    \bar n_A[1+\omega_{PA}(r_p)]
    +
    \bar n_B[1+\omega_{PB}(r_p)]
  },
  \label{eq:fA_given_P_projected}
\end{equation}
where the projected correlation with the {full neighbour sample
} satisfies
\begin{equation}
  1+\omega_{P{\rm g}}
  =
  \bar f_A[1+\omega_{PA}]
  +(1-\bar f_A)[1+\omega_{PB}] .
\end{equation}

We define
\begin{equation}
  \Delta f_A(r_p)
  =
  f_A(r_p|H)-f_A(r_p|L).
  \label{eq:def_deltaf_projected}
\end{equation}
Writing \(\Omega_{PY}=1+\omega_{PY}\), this is
\begin{equation}
\begin{aligned}
  \Delta f_A(r_p)
  &=
  \frac{\bar f_A \Omega_{HA}}
       {\bar f_A \Omega_{HA}+(1-\bar f_A)\Omega_{HB}}
  -
  \frac{\bar f_A \Omega_{LA}}
       {\bar f_A \Omega_{LA}+(1-\bar f_A)\Omega_{LB}} .
\end{aligned}
\label{eq:deltaf_fraction_projected}
\end{equation}
Thus \(\Delta f_A\) compares the neighbour mix around two primary classes.

In the weak-projected-correlation limit,
\begin{equation}
  \Delta f_A(r_p)
  \simeq
  \bar f_A(1-\bar f_A)
  \left[
    \omega_{HA}-\omega_{HB}-\omega_{LA}+\omega_{LB}
  \right],
  \label{eq:deltaf_linearized_omega}
\end{equation}
where all functions are evaluated at \(r_p\). This motivates the projected
cross-compensated statistic
\begin{equation}
  C_w(r_p)
  =
  w_{HA}(r_p)-w_{HB}(r_p)-w_{LA}(r_p)+w_{LB}(r_p),
  \label{eq:Cw_def}
\end{equation}
for which
\begin{equation}
  \Delta f_A(r_p)
  \simeq
  \bar f_A(1-\bar f_A)
  \frac{C_w(r_p)}{2\pi_{\rm max}} .
  \label{eq:deltaf_Cw_relation}
\end{equation}
The corresponding monopole is
\begin{equation}
  C_0(s)
  =
  \xi_{HA,0}(s)-\xi_{HB,0}(s)-\xi_{LA,0}(s)+\xi_{LB,0}(s).
  \label{eq:C0_def}
\end{equation}
{An important practical distinction is that $\Delta f_A$
can be measured directly from ratios of neighbour pair counts,
so the common random-catalogue normalization cancels.
By contrast, $C_w$ and $C_0$ are constructed from estimated
correlation functions and retain dependence on the random
catalogue. Their sensitivity to finite random sampling
has not been quantified here.}

We also consider {all-primary selection}, in which every galaxy
in the parent sample can act as a primary. This admits
satellite-primary pairs and provides a stress test of the
default central-primary results.

{We additionally define an auto-compensated statistic
using the same galaxy selection on both sides of each pair:}
\begin{equation}
  C_{w,\rm auto}(r_p)
  =
  w_{HH}(r_p)-2w_{HL}(r_p)+w_{LL}(r_p),
  \label{eq:Cw_auto_def}
\end{equation}
with monopole analogue
\begin{equation}
  C_{0,\rm auto}(s)
  =
  \xi_{HH,0}(s)-2\xi_{HL,0}(s)+\xi_{LL,0}(s).
  \label{eq:C0_auto_def}
\end{equation}
{Here $H$ and $L$ denote the high- and low-colour
populations within the chosen selection, either centrals
only or all galaxies. Unlike $C_w$, which uses the full
neighbour sample, $C_{w,\rm auto}$ restricts both members
of each pair to the selected population. Under all-primary
selection, $H=A$ and $L=B$, so
$C_{w,\rm auto}=C_w^{\rm all}$.}

Appendix~\ref{app:nulls} summarizes the additional compensated combinations
used as consistency tests.

\subsection{Matter templates and response amplitudes}
\label{sec:linear_templates}

We compare the measured statistics with linear-matter templates,
{focusing on their shapes and fitting an independent amplitude
for each statistic.} For projected statistics the formally matched
template is
\begin{equation}
  M_w(r_p)=A_w\,w_{\rm mm}^{\rm lin}(r_p),
  \ 
  w_{\rm mm}^{\rm lin}(r_p)
  =
  2\int_0^{\pi_{\rm max}}
  \xi_{\rm mm}^{\rm lin}\!\left(\sqrt{r_p^2+\pi^2}\right)d\pi .
  \label{eq:wmm_template}
\end{equation}
For monopole statistics we use
\begin{equation}
  M_\xi(s)=A_\xi\,\xi_{\rm mm}^{\rm lin}(s).
  \label{eq:ximm_template}
\end{equation}
The fitted amplitudes absorb the galaxy--halo connection, the strength of the
binary split, and the response of the selected galaxy property to the density
field.

For simulations we evaluate \(\xi_{\rm mm}^{\rm lin}\) using the corresponding
simulation cosmology. For SDSS and DESI we keep the catalogue coordinates
fixed and use the fiducial Planck-like template. {We define
\(\Omega_{\rm m}^{\rm shape}\) as an effective parameter that varies
the linear template's broad-band shape and its AP remapping,
with other cosmological parameters fixed. It is an exploratory
shape constraint, not a full BAO, AP, or RSD measurement
(Appendix~\ref{app:omega_shape}).}

For any statistic \(S(r)\), with \(r=r_p\) or \(s\), and any chosen matter
template \(T_{\rm mm}^{\rm lin}(r)\), we define
\begin{equation}
  S(r)={\cal A}_{S,T}(r)\,T_{\rm mm}^{\rm lin}(r).
  \label{eq:Aofr_general_def}
\end{equation}
A single-amplitude fit assumes
\begin{equation}
  S(r)\simeq A_{\rm fit}\,T_{\rm mm}^{\rm lin}(r),
\end{equation}
so the diagnostic ratio is
\begin{equation}
  \frac{S(r)}
       {A_{\rm fit}T_{\rm mm}^{\rm lin}(r)} .
  \label{eq:Aofr_ratio_plot}
\end{equation}
A flat ratio indicates that the statistic has the shape of the chosen matter
template over the fitted range.

Projected compensated statistics are also compared with the diagnostic template
\begin{equation}
  M_{\xi,p}(r_p)=A_{\xi,p}\,\xi_{\rm mm}^{\rm lin}(r_p).
  \label{eq:ximm_projected_null_template}
\end{equation}
This is not the formal projection of a two-point correlation. It tests whether
the residual projected signal follows the local, unprojected matter-correlation
shape instead of the broadened projected template \(w_{\rm mm}^{\rm lin}\).

\subsection{Nonlinear mode cancellation}
\label{sec:common_mode}

We interpret the compensated statistics {using a phenomenological
decomposition of} each correlation
into the linear matter mode and additional scale-dependent clustering modes labelled $n$:
\begin{equation}
\xi_{ab}(r)
=
c_{ab}^{(0)}\,\xi_{\rm mm}^{\rm lin}(r)
+
\sum_{n>0}c_{ab}^{(n)}\,G_n(r).
\label{eq:mode_decomp_generic}
\end{equation}
{The $G_n(r)$ represent generic contributions to galaxy
clustering beyond the linear matter term, including nonlinear
density evolution, galaxy bias, and halo occupation effects.
Their functional forms and number are unspecified, as the
expansion is schematic.}

For the four-term compensated statistic,
\begin{equation}
C(r)
=
\Delta c^{(0)}\,\xi_{\rm mm}^{\rm lin}(r)
+
\sum_{n>0}\Delta c^{(n)}\,G_n(r),
\label{eq:C_mode_decomp}
\end{equation}
where each \(\Delta c^{(n)}\) is the corresponding four-term difference of
mode coefficients. Compensation is therefore a projection in population
space: it does not explicitly select the linear mode, but suppresses any mode
whose response is similar among the combined samples,
\(\Delta c^{(n)}\simeq0\). Our empirical hypothesis is that
assembly-sensitive splits suppress several nonlinear modes while
\(\Delta c^{(0)}\ne0\), allowing the linear matter mode to dominate the
residual. {Assembly-dependent contributions may therefore survive
in the linear response while their scale-dependent components are
partially suppressed.} The {all-primary selection tests the
robustness of this cancellation to enhanced} halo-scale and satellite
contributions.

\subsection{Minimum scale of linear-template consistency}
\label{sec:rmin}

{We define $r_{\min}$ as the smallest separation down to which
a statistic is consistently described by a single-amplitude
linear matter template. We determine it by fitting $X(r)=A\,T_{\rm mm}^{\rm  lin}(r)$
over successive ranges $[r_{\min},30]\,h^{-1}{\rm Mpc}$,
using the full jackknife covariance. A fit is accepted if
its chi-square goodness-of-fit probability satisfies $p\geq0.05$
and no individual bin deviates from the model by more than
twice its diagonal jackknife uncertainty. We require at least
four bins and define $r_{\min}$ as the smallest lower scale cut
for which these conditions hold for all larger cuts.}

{This criterion is intended as a first operational
estimate of the minimum scale of linear-template consistency.
Its sensitivity to alternative goodness-of-fit thresholds
has not been explored here.}

\subsection{Effective projected kernels and Kaiser cancellation}
\label{sec:effective_kernels}

We characterise the residual clustering response through an
effective line-of-sight kernel. For the cross-compensated
statistic, we define
\begin{equation}
  \widehat W_C(r_p,\pi)
  =
  \frac{C(r_p,\pi)}
       {A_C\,\xi_{\rm mm}^{\rm lin}(s)},
  \label{eq:WC_hat_def}
\end{equation}
where $s=\sqrt{r_p^2+\pi^2}$ and $A_C$ is the fitted
amplitude. The projected statistic is then
\begin{equation}
  C_w(r_p)
  =
  2A_C\int_0^{\pi_{\rm max}}
  \widehat W_C(r_p,\pi)
  \xi_{\rm mm}^{\rm lin}(s)\,d\pi.
  \label{eq:Cw_kernel_projection}
\end{equation}
A constant kernel recovers the projected linear matter
template. A response concentrated at small line-of-sight
separations can instead produce a projected statistic
whose shape more closely resembles
$\xi_{\rm mm}^{\rm lin}(r_p)$.
The corresponding derivation for the exact, unprojected
conformity statistic is given in Appendix~\ref{app:kernels}.

{For monopoles, recovery of the linear matter correlation-function shape
requires an approximately scale-independent angularly averaged response.
The conformity statistic provides an additional simplification in redshift
space. In linear theory, the Kaiser cross-power spectrum of populations
$i$ and $j$ is
\begin{equation}
P_{ij}^{s}(k,\mu)
= (b_i+f\mu^2)(b_j+f\mu^2)P_{\rm mm}^{\rm lin}(k),
\end{equation}
where $f$ is the logarithmic growth rate and $\mu$ is the cosine of
the angle between the wavevector and the line of sight.}

{On sufficiently large scales (small $k$ values), linearising the conformity fractions
gives a four-term combination of correlations. Its Fourier transform is
\begin{align}
\widetilde{\Delta f}_X^{s}(k,\mu)
&\simeq \bar f_X
\left[P_{HX}^{s}-P_{HN}^{s}-P_{LX}^{s}+P_{LN}^{s}\right]\nonumber\\
&=\bar f_X(b_H-b_L)(b_X-b_N)P_{\rm mm}^{\rm lin}(k),
\end{align}
where $N$ denotes the full neighbour sample. Thus the leading Kaiser
terms cancel, leaving an isotropic linear response. Consequently,
\begin{equation}
\Delta f_{X,2}^{\rm lin}(s)=0,
\end{equation}
while the monopole retains the shape of $\xi_{\rm mm}^{\rm lin}(s)$,
with an amplitude set by the population bias differences.}

Thus projected statistics can recover an
\(\xi_{\rm mm}^{\rm lin}\)-like shape through a localized line-of-sight
kernel, whereas monopoles {require a nearly scale-independent
response; the cancellation of leading Kaiser terms additionally
suppresses their redshift-space anisotropy.}
We expect the ratio statistic \(\Delta f_A\) to
satisfy {the scale-independence condition} more readily than
the linear combination \(C\).

\section{Data samples}
\label{sec:data}

We use the DESI Bright Galaxy Survey (BGS) as our main observational sample
and the Sloan Digital Sky Survey Main Galaxy Sample (SDSS MGS) as a
low-redshift comparison. Both are used to construct rest-frame-\(r\)-band
fixed-number-density samples. SDSS MGS has \(r<17.77\)
\citep{Strauss2002MGS,Zehavi2011}, while DESI BGS extends approximately to
\(r<19.5\) \citep{RuizMacias2020BGS,DESI2024SampleDefinitions}.

\subsection{DESI BGS, SDSS MGS, and central-primary catalogues}
\label{sec:data_surveys}

For DESI we use the BGS clustering catalogues from the DESI large-scale
structure data products, together with the corresponding random catalogues and
standard spectroscopic weights
\citep{Ross2024DESILSS,DESI2024SampleDefinitions}. Rest-frame colours and
absolute magnitudes are taken from DESI value-added products. Throughout this
work, DESI BGS refers to the combined north and south galactic caps (NGC and
SGC) footprint. {For central-primary selection, we identify
centrals using} the DESI DR9/Y1 halo-based group-finder catalogue, part of
the DR1 DESI data products, matched to the BGS clustering catalogues through
the imaging identifiers \texttt{RELEASE}, \texttt{BRICKID}, and
\texttt{OBJID} decoded from \texttt{TARGETID}. We identify central galaxies
as group members with \texttt{RANK}=0.

For SDSS we use the Main Galaxy Sample, with the standard angular mask,
spectroscopic completeness information, and survey weights
\citep{Strauss2002MGS,Zehavi2011}. Central galaxies are identified as the
brightest group members using the halo-based SDSS group catalogue of
\citet{Yang2005GroupFinder,Yang2007SDSSGroups}. For both surveys,
{central-primary selection restricts only the primary side of the
pair counts to identified centrals; the neighbour sample remains the
full fixed-number-density parent sample, including satellites.}
This preserves the usual conformity interpretation, in which the neighbour
population is compared around two classes of primary central galaxies.

All galaxy selections use SDSS \(g\) and \(r\) bands, \(k\)-corrected to
\(z=0.1\), following the standard convention for low-redshift SDSS clustering
analyses \citep{BlantonRoweis2007,Zehavi2011}. We denote the corresponding
absolute magnitude and colour by \(^{0.1}M_r\) and \(^{0.1}(g-r)\). For each
survey we construct {parent samples at fixed number densities}
\begin{equation}
n = 0.003,\; 0.010,\; 0.020\,h^3{\rm Mpc}^{-3},
\end{equation}
where supported by the survey depth and redshift range. The highest-density
sample, \(n=0.020\,h^3{\rm Mpc}^{-3}\), is included because
\citet{PadillaLacernaPaz2026Letter} showed that the linear-response behaviour
of conformity becomes cleaner at higher galaxy number density. The magnitude
and redshift limits ensure volume completeness at each number density
(Table~\ref{tab:data_samples}). Increasing number density requires
fainter thresholds and lower maximum redshifts, so the samples differ in both
statistical precision and intrinsic galaxy population.

\begin{table*}
\centering
\caption{
Galaxy samples used in the DESI BGS and SDSS MGS measurements.  Absolute
magnitudes and colours are rest-frame SDSS quantities \(k\)-corrected to
\(z=0.1\).  \(N_{\rm gal}\) is the number of galaxies in the parent sample,
\(f_{\rm cen}\) is the fraction identified as centrals, and
\(^{0.1}(g-r)_{\rm cut}\) is the median colour threshold used for the
equal-number split.
}
\label{tab:data_samples}
\begin{tabular}{lcccccccc}
\hline\hline
Survey
& \(n\)
& \(z_{\rm min}\)
& \(z_{\rm max}\)
& \(^{0.1}M_r\) cut
& apparent \(r\) limit
& \(N_{\rm gal}\)
& \(f_{\rm cen}\)
& \(^{0.1}(g-r)_{\rm cut}\)
\\
&
\([h^3{\rm Mpc}^{-3}]\)
&
&
&
&
&
&
&
\\
\hline
SDSS MGS
& \(0.003\)
& 0.020
& 0.100
& \(\leq -20.505\)
& \(r<17.77\)
& \(52\,906\)
& 0.790
& 0.894
\\
DESI BGS
& \(0.003\)
& 0.020
& 0.200
& \(\leq -20.156\)
& \(r<19.50\)
& \(451\,332\)
& 0.725
& 0.867
\\
SDSS MGS
& \(0.010\)
& 0.020
& 0.093
& \(\leq -19.602\)
& \(r<17.77\)
& \(142\,213\)
& 0.728
& 0.845
\\
DESI BGS
& \(0.010\)
& 0.020
& 0.130
& \(\leq -18.656\)
& \(r<19.50\)
& \(477\,627\)
& 0.670
& 0.758
\\
SDSS MGS
& 0.02
& 0.020
& 0.058
& \(\leq -18.532\)
& \(r<17.77\)
& \(106\,152\)
& 0.687
& 0.766
\\
DESI BGS
& 0.02
& 0.010
& 0.080
& \(\leq -17.242\)
& \(r<19.50\)
& \(251\,324\)
& 0.623
& 0.627
\\
\hline
\end{tabular}
\end{table*}

\begin{figure}
    \centering
    \includegraphics[width=\linewidth]{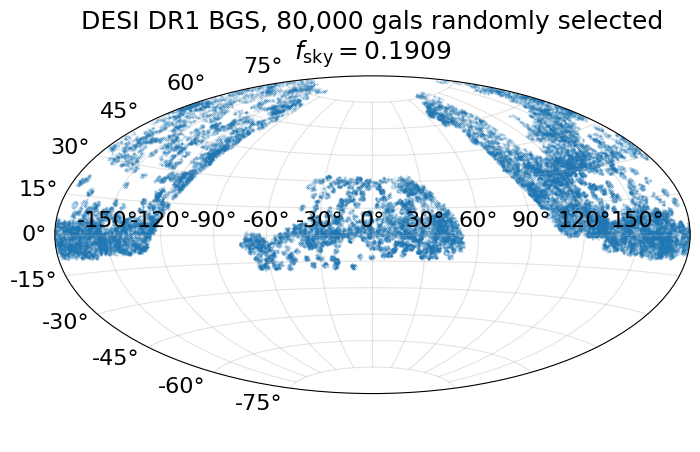}
    \caption{
    Angular footprint of the DESI DR1 BGS sample used for our measurements.  The
    points show a sparse subsample of BGS galaxies.
    }
    \label{fig:desi_bgs_footprint}
\end{figure}

The DESI DR1 BGS angular footprint contains both contiguous regions and
isolated pointings, as shown in Fig.~\ref{fig:desi_bgs_footprint}. This matters
for projected measurements because the DESI field of view has a diameter of
\(3.2^\circ\), corresponding to transverse comoving scales of approximately
\(16\), \(21\), and \(32\,h^{-1}{\rm Mpc}\) at \(z=0.1\), \(0.13\), and
\(0.2\), respectively. These scales overlap the range used for the projected
compensated statistics. {We account for the survey window using the
DESI clustering random catalogues, while noting that the sensitivity of
compensated measurements to random-catalogue sampling has not been
independently assessed.}

\subsection{Galaxy splits}
\label{sec:data_splits}

For each survey and number-density {parent sample}, we split galaxies
into equal-number red and blue subsamples using the median rest-frame colour
\(^{0.1}(g-r)\) of the selected parent sample:
\begin{equation}
{\rm red}:
\quad
{}^{0.1}(g-r) > {}^{0.1}(g-r)_{\rm med},
\end{equation}
and
\begin{equation}
{\rm blue}:
\quad
{}^{0.1}(g-r) < {}^{0.1}(g-r)_{\rm med}.
\end{equation}
The median is computed separately for each survey and each number-density
sample (see Table~\ref{tab:data_samples}). {This single threshold
defines both the red and blue neighbour populations ($A$ and $B$)
and the corresponding red and blue primary populations ($H$ and $L$).
For central-primary selection, only identified centrals enter $H$ and $L$;
for all-primary selection, all galaxies can enter these populations.
In both cases, neighbours are drawn from the full parent sample.}

This rank-based split avoids imposing a common absolute colour threshold on
DESI and SDSS, whose photometric inputs, selection functions, and redshift
distributions differ. Colour is an assembly-sensitive galaxy property:
simulation and empirical studies show that galaxy assembly bias and two-halo
conformity depend on colour, star-formation activity, and related secondary
properties at approximately fixed halo mass
\citep{Croton2007AssemblyBias,Wang2013AssemblyBias,
Hearin2015Beyond,Zentner2014AssemblyBias,Hearin2016DecoratedHOD,
Lacerna2018Conformity,MonteroDorta2020,PadillaLacernaPaz2026Letter}.

We use the same binary split for \(\Delta f_{\rm color}\), \(C_w\),
and \(G_3^w\), the latter defined in Appendix~\ref{app:nulls}.
We do not use star-formation-rate or spectral-class splits in this paper.

\subsection{Random catalogues, weights, and covariance matrices}
\label{sec:data_randoms}

The DESI measurements presented here use the official BGS clustering random catalogues, which
encode the angular and radial selection function of the data
\citep{Ross2024DESILSS,DESI2024SampleDefinitions}. For DESI we use the
standard large-scale-structure weights, including targeting, imaging,
spectroscopic, and redshift-success corrections. For SDSS MGS we use the
standard survey weights and angular mask associated with the spectroscopic
sample \citep{Strauss2002MGS,Zehavi2011}. Pair counts use the product of the
weights of the two objects in each pair, with the same convention for
data--data ($DD$), data--random ($DR$), and random--random ($RR$) terms.
{We estimate the two-point correlation functions using the
Landy--Szalay estimator,
$\xi=(DD-2DR+RR)/RR$, with the corresponding cross-correlation
form for distinct galaxy populations. All pair counts are
appropriately normalized and use the same survey random
catalogue. The conformity fractions are instead measured
directly from ratios of neighbour pair counts.}

Our baseline DESI measurements use one clustering random catalogue. We
therefore base the precision-dependent part of the analysis on ratios and
matched combinations of auto- and cross-correlations measured with the same
selection function, for which {common random-catalogue contributions
are expected to be partially suppressed}. Statistics that depend more directly
on data--random and random--random counts, in particular \(C_0\), are used
mainly as diagnostics.

All measurements use \(64\) jackknife regions defined over the survey
footprint. The covariance matrix of each statistic is estimated as
\begin{equation}
{\rm Cov}_{ij}
=
\frac{N_{\rm JK}-1}{N_{\rm JK}}
\sum_{k=1}^{N_{\rm JK}}
\left[
S_i^{(k)}-\bar S_i
\right]
\left[
S_j^{(k)}-\bar S_j
\right],
\end{equation}
where \(S_i^{(k)}\) is the measurement in radial bin \(i\) after removing
jackknife region \(k\), \(\bar S_i\) is the mean over jackknife samples, and
\(N_{\rm JK}=64\).

\subsection{Pair-count measurements}
\label{sec:data_paircounts}

For each survey, density, and colour split, we measure
central-primary and all-primary correlations using identical
neighbour samples, weights, jackknife regions, and binning.
The former cross-correlates red and blue centrals with the
full red and blue populations; the latter uses the full
populations on both sides of each pair.

{Projected statistics use $(r_p,\pi)$ counts integrated to
$\pi_{\max}=80\,h^{-1}{\rm Mpc}$; monopoles and quadrupoles
use $(s,\mu)$ counts. We adopt 30 logarithmic separation
bins over $0.5$--$80\,h^{-1}{\rm Mpc}$ and 80 uniform
bins in both $\mu\in[0,1]$ and $\pi\in[0,80]\,h^{-1}{\rm Mpc}$.}

\section{Simulated galaxy catalogues}
\label{sec:sims}

\begin{table*}
\caption{Galaxy samples used for comparison with the galaxy-formation models. For MTNG and FLAMINGO, samples are selected by $r$-band absolute magnitude rank to obtain the target number density $M_{r,\max}$. For MDPL2--SAG, galaxies are selected using an absolute-magnitude limit. For computational tractability, at most $5\times10^{6}$ MDPL2--SAG galaxies are retained for the clustering calculations.}
\label{tab:simulation_samples}
\centering
\begin{tabular}{lcccccc}
\hline\hline
Model
& $L_{\rm box}$
& $n$
& $N_{\rm gal}$
& $(g-r)$ threshold
& limit $M_r$ \\
& [$h^{-1}\,{\rm Mpc}$]
& [$h^{3}\,{\rm Mpc}^{-3}$]
&
&
&
& \\
\hline
MTNG
& 500
& 0.003
& 375\,000
& 0.534
& -21.69\\

FLAMINGO
& 681
& 0.003
& 947\,464
& 0.7012
& -21.39 \\

MDPL2--SAG
& 1000
& 0.003
& 3\,000\,000
& 0.5402
&-21.08 \\
\hline

MTNG
& 500
& 0.010
& 1\,250\,000
& 0.424
& --19.98 \\

FLAMINGO
& 681
& 0.010
& 3\,158\,212
& 0.6881 
& -20.45 \\

MDPL2--SAG
& 1000
& 0.010
& 10\,000\,000$^{a}$
& 0.4776
& -20.24 \\
\hline

MTNG
& 500
& 0.0204
& 2\,550\,000
& 0.446 
& -18.16\\

FLAMINGO
& 681
& 0.0204
& 6\,448\,242
& 0.6906
& -18.80 \\

MDPL2--SAG
& 1000
& 0.0204
& 20\,080\,000$^{a}$
& 0.4001
& -19.20\\
\hline
\end{tabular}

\begin{flushleft}
$^{a}$ For the clustering calculations a maximum of randomly selected $5\,000\,000$ galaxies were used for these samples.
\end{flushleft}
\end{table*}

We use MillenniumTNG, FLAMINGO, and MDPL2--SAG as independent testbeds for the
compensated statistics.  These simulations are not calibrated to reproduce the
DESI or SDSS
colour-dependent clustering. They test recovery of the known input
matter-clustering shape across hydrodynamical and semi-analytic
galaxy-formation prescriptions, with MTNG providing the closest continuation
of the TNG300 experiment of \citet{PadillaLacernaPaz2026Letter}.

\subsection{Simulation matching}
\label{sec:sims_matching}

The simulated samples are matched to the observational number-density and
colour-rank selections. For each simulation and target number density, we rank
galaxies by rest-frame \(r\)-band absolute magnitude and select the brightest
objects required to match the desired comoving density. We use
\(n=0.003\), \(0.010\,h^3{\rm Mpc}^{-3}\), and the
higher-density sample with
\(n/[h^3{\rm Mpc}^{-3}]=0.02\).

Red and blue galaxies are defined by a global median split in rest-frame
\(g-r\) colour within each selected parent sample, matching the rank-based
observational definition (see Sect. \ref{sec:data_splits}). {For central-primary selection, we use
the central/satellite information available in each catalogue to restrict
the primary sample to central galaxies. The neighbour sample remains the
full fixed-number-density parent sample, including centrals and satellites.
We also consider all-primary selection, in which every selected galaxy
can act as a primary. The same median colour threshold is used for both
primary selections and for the neighbour populations.}

All simulated catalogues are analysed in real space and in
redshift space. Redshift-space positions are obtained by displacing galaxies
along a Cartesian line of sight using their peculiar velocities. We measure
the same projected and redshift-space statistics used for the observations:
\((r_p,\pi)\) statistics integrated to \(\pi_{\rm max}\), and monopoles from
\((s,\mu)\) pair counts integrated over \(\mu\).

\subsection{Galaxy-formation simulations}

\begin{figure*}
  \centering
  \includegraphics[width=\textwidth]{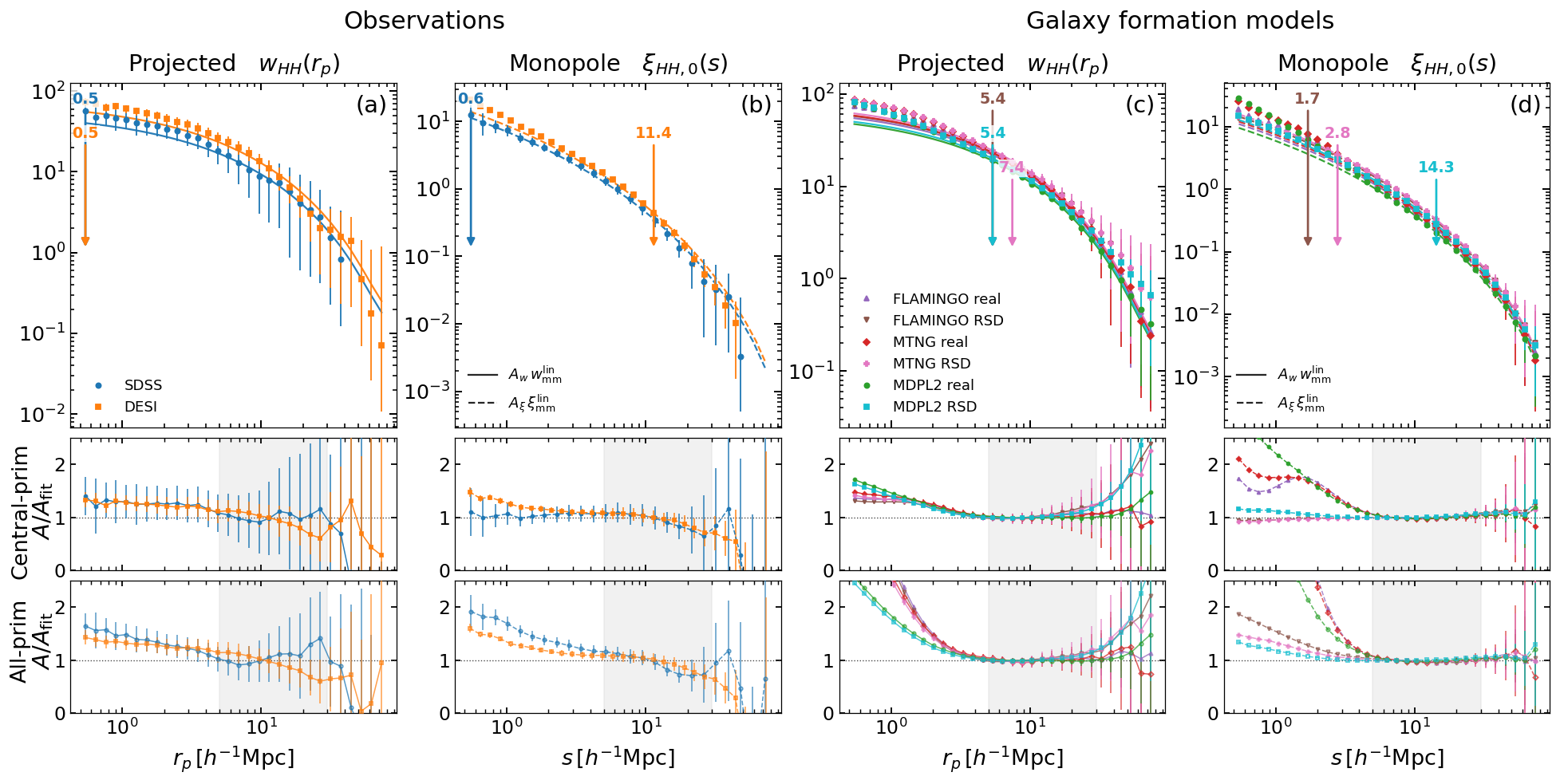}
\caption{
Ordinary high-colour clustering in the observations (left two columns) and
galaxy-formation models (right two columns). The first and third columns show
the projected statistic \(w_{HH}(r_p)\), while the second and fourth columns
show the redshift-space monopole \(\xi_{HH,0}(s)\). The upper panels show the
measured statistics together with the best-fitting linear matter templates.
The middle and lower panels show the response ratios for
{central-primary and all-primary selections}, respectively.
The observational results correspond to
SDSS MGS and DESI BGS, while the simulations include FLAMINGO, MTNG, and
MDPL2--SAG in real and redshift space. Shaded regions indicate the fiducial
fitting ranges. {Arrows indicate the minimum scale of
linear-template consistency, $r_{\min}$, defined in
Sect.~\ref{sec:rmin}.}
}
  \label{fig:ordinary_clustering}
\end{figure*}

\begin{figure*}
  \centering
  \includegraphics[width=\textwidth]{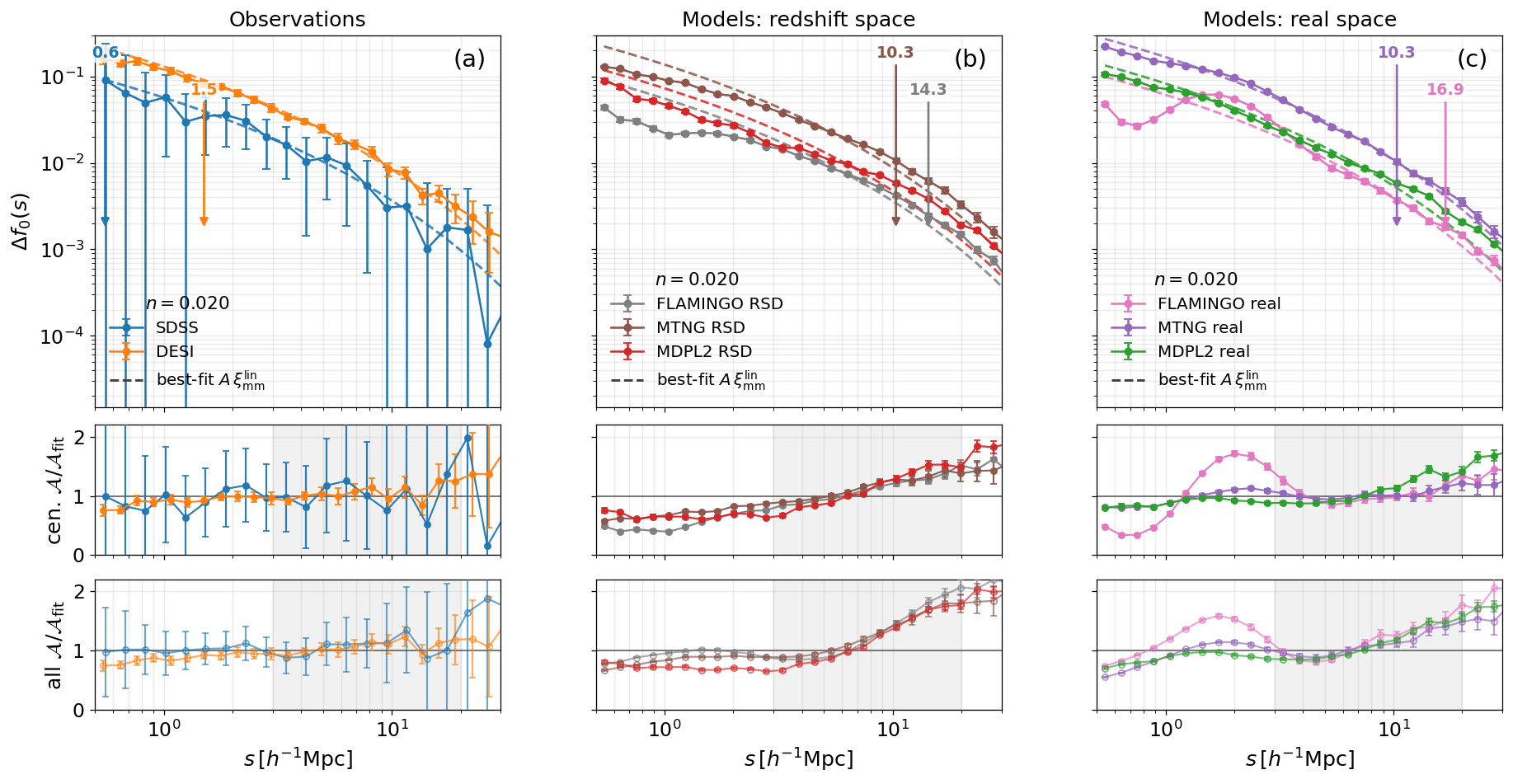}
\caption{
Conformity monopole response. The left column shows SDSS MGS and DESI BGS,
the middle column shows simulated catalogues in redshift space, and the right
column shows simulated catalogues in real space. Upper panels show
\(\Delta f_0(s)\) and the best-fitting \(A_\xi\xi_{\rm mm}^{\rm lin}(s)\)
templates. Middle panels show {response ratios for central-primary
selection}, and lower panels show {those for all-primary selection}.
Shaded regions indicate the fiducial fitting range;
{arrows indicate $r_{\min}$ as defined in
Sect.~\ref{sec:rmin}.}
}
  \label{fig:deltaf0_response}
\end{figure*}
\begin{figure*}
  \centering
  \includegraphics[width=0.95\textwidth]{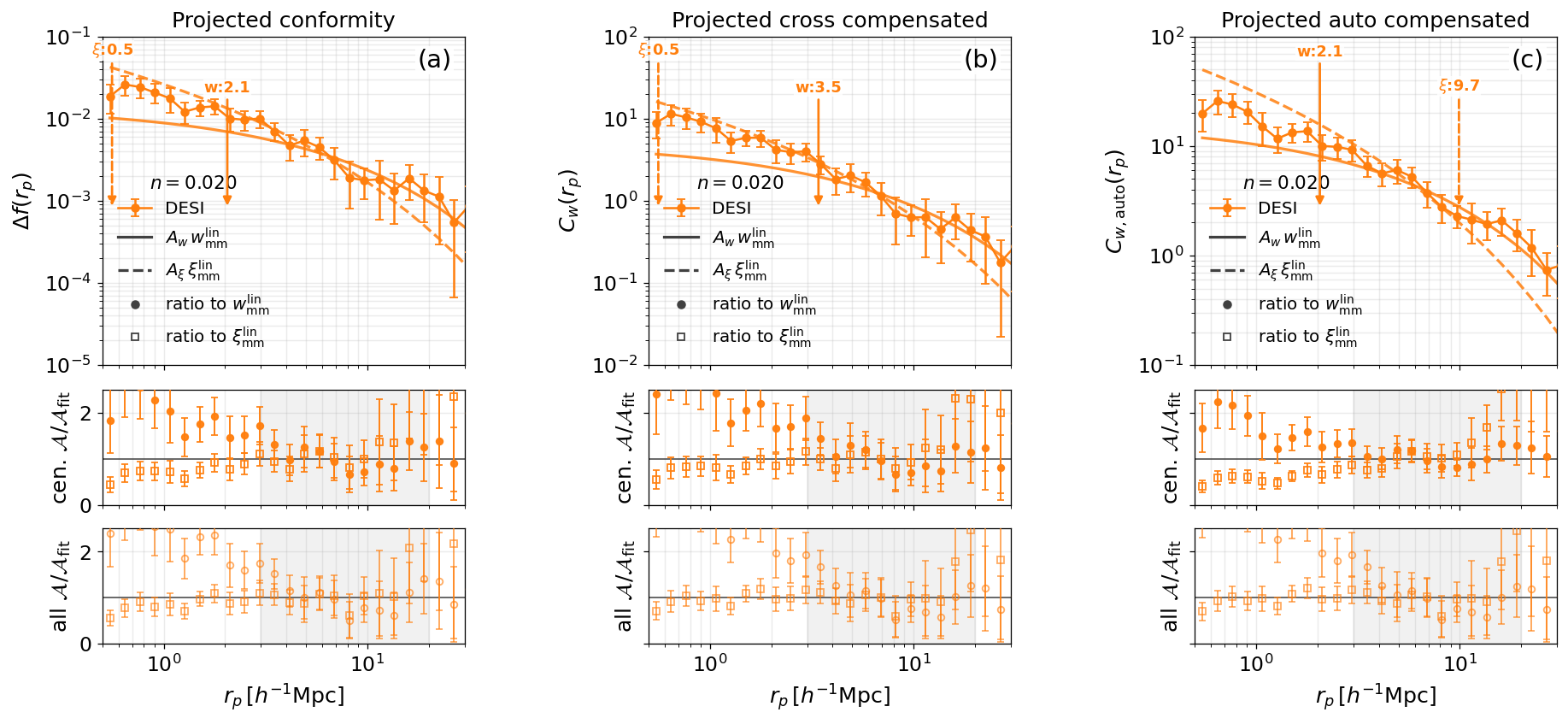}

  \includegraphics[width=0.95\textwidth]{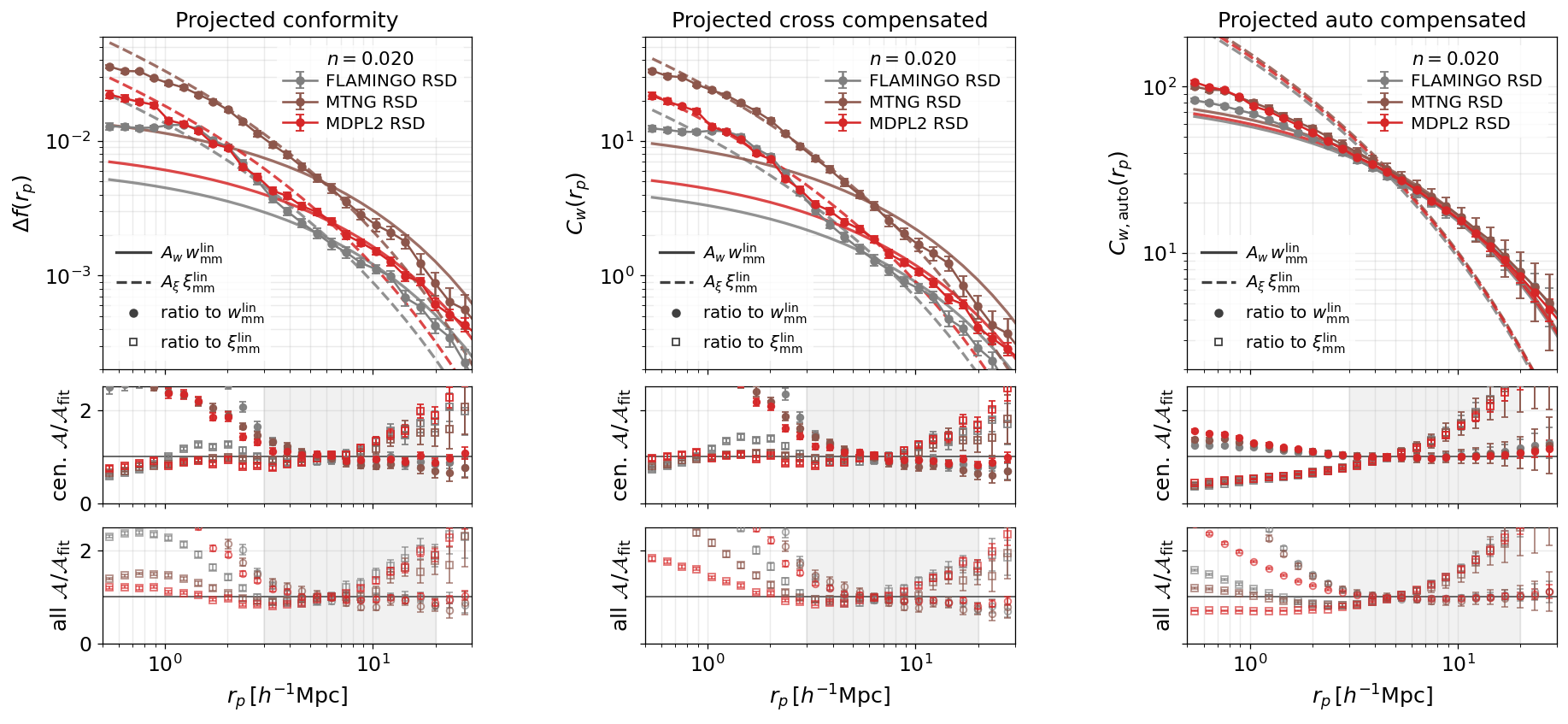}

\caption{
Projected conformity and compensated clustering statistics. The upper set of
panels shows DESI BGS measurements, while the lower set shows the simulated
catalogues in redshift space. From left to right, the columns show the
projected conformity statistic \(\Delta f(r_p)\), the projected
{cross-compensated} statistic \(C_w(r_p)\), and the projected
{auto-compensated} statistic \(C_{w,\mathrm{auto}}(r_p)\).
The upper row in each set shows the measured statistics together with the
best-fitting projected linear matter templates
\(A_w w_{\rm mm}^{\rm lin}(r_p)\) (solid) and
\(A_\xi \xi_{\rm mm}^{\rm lin}(r_p)\) (dashed).
The middle and lower rows show the corresponding response ratios for the
{central-primary and all-primary selections}, respectively.
Filled circles denote ratios to
\(w_{\rm mm}^{\rm lin}(r_p)\), and open squares denote ratios to
\(\xi_{\rm mm}^{\rm lin}(r_p)\).
Shaded regions indicate the fiducial fitting range;
{solid and dashed arrows, shown only in the top panel, indicate
$r_{\min}$ for the projected and unprojected linear matter templates,
respectively.}
}
  \label{fig:projected_compensated}
\end{figure*}


\begin{figure*}
  \centering
  \includegraphics[width=0.8\textwidth]{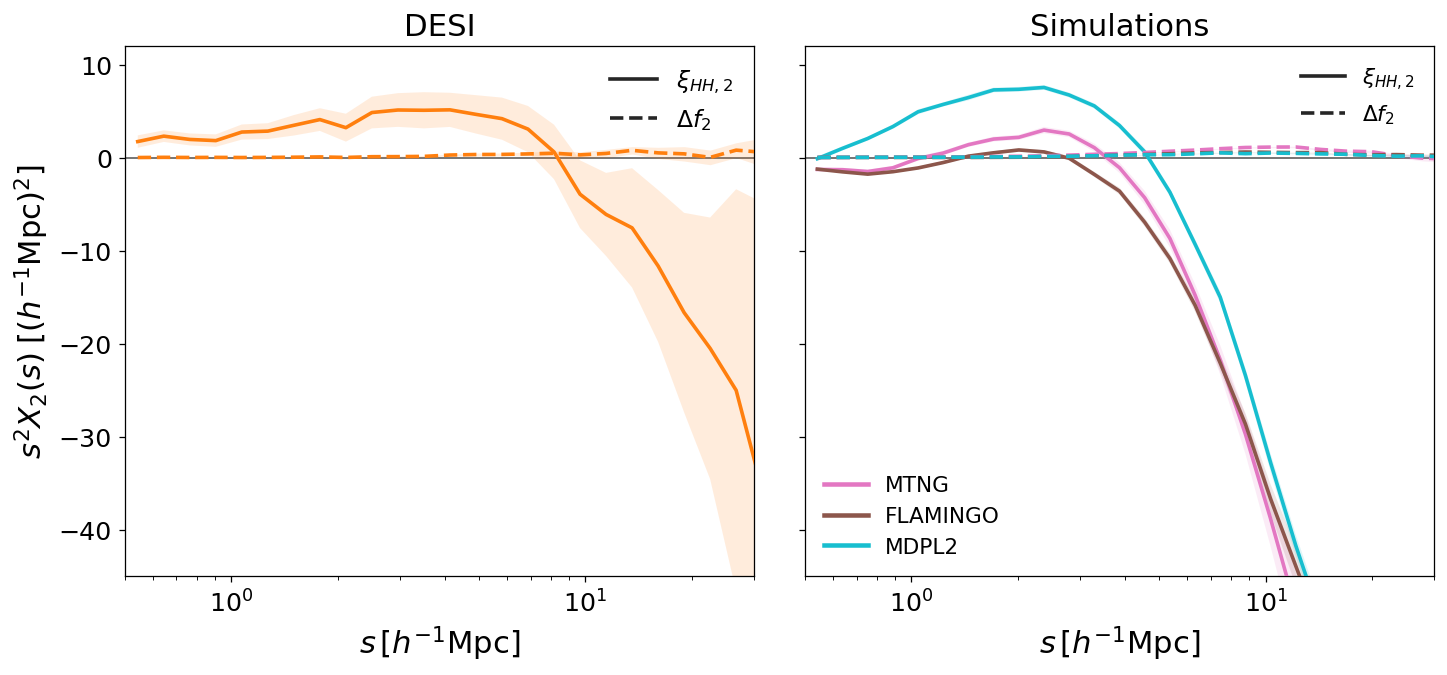}
\caption{
Redshift-space correlation function (solid) and conformity (dashed)
quadrupoles {measured with central-primary selection}.
The left panel shows results for DESI BGS.
The right panel {shows results for} MDPL2--SAG,
MTNG, and FLAMINGO.
}
  \label{fig:quadrupoles}
\end{figure*}

\begin{figure*}
  \centering
  \includegraphics[width=0.98\textwidth]{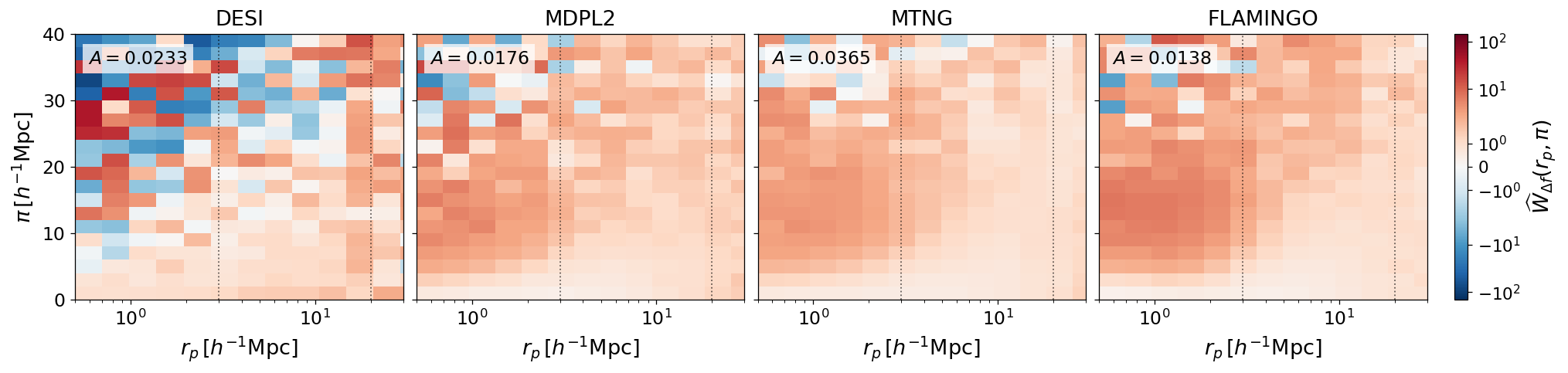}
\caption{
Effective line-of-sight kernels for the projected conformity statistic
\(\Delta f(r_p)\). Columns correspond to DESI, MDPL2--SAG, MTNG, and
FLAMINGO, {measured with central-primary selection
(similar results are obtained with all-primary selection)}.
The colour scale shows the normalized effective
kernel
\(\widehat{W}_{\Delta f}(r_p,\pi)
=\Delta f(r_p,\pi)/[A_{\Delta f}\xi_{\rm mm}^{\rm lin}(s)]\),
with \(s=\sqrt{r_p^2+\pi^2}\). The normalization amplitude
\(A_{\Delta f}\) used in each panel is indicated in the upper-left corner.
}
  \label{fig:effective_projection_kernels}
\end{figure*}

We compare the observations with three galaxy-formation models. MillenniumTNG
(MTNG) is a hydrodynamical simulation combining the IllustrisTNG
galaxy-formation model with a volume of
\((500\,h^{-1}{\rm Mpc})^3\)
\citep{Pakmor2023MTNGHydro,Bose2023MTNGClustering}. It follows dark matter,
gas, stars, and black holes, including subgrid prescriptions for cooling, star
formation, feedback, chemical enrichment, and black-hole growth. We use its
\(z=0\) galaxy catalogue, including positions, velocities, stellar masses, and
rest-frame SDSS photometry. MTNG adopts
\(\Omega_m=0.3089\), \(\Omega_\Lambda=0.6911\), \(\Omega_b=0.0486\),
\(h=0.6774\), \(\sigma_8=0.8159\), and \(n_s=0.9667\).

FLAMINGO is a suite of large-volume hydrodynamical simulations developed for
large-scale-structure, cluster, and baryonic-effect studies
\citep{Schaye2023FLAMINGO,Kugel2023FLAMINGOCalibration}. We use the \(z=0\)
catalogue from the \(L1\_m8\) run, which has a \(1\,{\rm Gpc}\) box and provides
positions, velocities, stellar masses, and broad-band luminosities within a
50 kpc aperture, from which we construct rest-frame \(M_r\) and \(g-r\).
Its fiducial D3A cosmology has
\(\Omega_m=0.306\), \(\Omega_\Lambda=0.694\), \(\Omega_b=0.0486\),
\(h=0.681\), \(\sigma_8=0.807\), and \(n_s=0.967\).

Finally, MDPL2 is a dark-matter-only simulation with volume
\((1\,h^{-1}{\rm Gpc})^3\), populated with galaxies using the SAG
semi-analytic model
\citep{Klypin2016MultiDark,Planck2013Cosmo,Cora2018SAG,
Knebe2018MultiDarkGalaxies}. SAG follows gas cooling, star formation,
feedback, chemical enrichment, black-hole growth, AGN feedback, and satellite
environmental processes, and provides SDSS broad-band magnitudes. The
MultiDark--Planck cosmology has
\(\Omega_m=0.307\), \(\Omega_\Lambda=0.693\), \(\Omega_b=0.048\),
\(h=0.678\), \(\sigma_8=0.829\), and \(n_s=0.96\).

For each simulation, we compute \(\xi_{\rm mm}^{\rm lin}\) and
\(w_{\rm mm}^{\rm lin}\) using its corresponding cosmology.
The resulting samples are summarized in
Table~\ref{tab:simulation_samples}. Their absolute photometric limits
differ because of their stellar-population, dust, and galaxy-formation
prescriptions, so neither the median \(g-r\) colour nor the sampled \(M_r\)
range is expected to agree exactly. We therefore define red and blue
subsamples relative to each sample's median \(g-r\). The listed
\(M_{r,\max}\) is the faintest selected galaxy.

\section{Compensated statistics}
\label{sec:results}

\subsection{Ordinary high-colour clustering}
\label{sec:ordinary_clustering}

Before analysing conformity and compensated statistics, we examine
the ordinary two-point clustering of the high-colour (``red'')
population, denoted by \(H\).

{Under central-primary selection}, \(w_{HA}(r_p)\) and
\(\xi_{HA,0}(s)\) cross-correlate high-colour centrals with
neighbours drawn from the full parent sample \(A\).
{Under all-primary selection}, the corresponding
\(HH\) measurements reduce to the usual high-colour
auto-correlation. These measurements provide the baseline
for the compensated statistics.

Figure~\ref{fig:ordinary_clustering} compares the ordinary clustering
measurements with linear matter templates. The projected correlations are fitted
with \(A_w w_{\rm mm}^{\rm lin}(r_p)\), while the monopoles are fitted with
\(A_\xi \xi_{\rm mm}^{\rm lin}(s)\). In each column, the upper panel shows the
measured statistic and its fitted template, the middle panel shows the response
ratio for {central-primary selection}, and the lower panel shows the same
ratio for {all-primary selection}:
\begin{equation}
  \frac{{\cal A}(r)}{A_{\rm fit}}
  =
  \frac{S(r)}
       {A_{\rm fit}T_{\rm mm}^{\rm lin}(r)} ,
\end{equation}
where \(S(r)\) denotes the measured statistic, \(T_{\rm mm}^{\rm lin}(r)\) is the
corresponding linear matter template, and \(r\) is either \(r_p\) or \(s\).
{We adopt a fitting range of $5$--$30\,h^{-1}{\rm Mpc}$
to characterize departures from the linear template
across intermediate and large scales.}

The ordinary correlations follow the linear matter templates on large scales
but depart from them toward intermediate scales, {as indicated by the
$r_{\min}$ values marked by vertical arrows (Sect.~\ref{sec:rmin}).}
{Their response ratios show corresponding scale dependence in both
observations and simulations.}
{DESI exhibits a larger clustering amplitude than SDSS
in these measurements. Differences in galaxy selection (see Table \ref{tab:data_samples}) may
contribute to this offset, while the smaller SDSS volume
increases the uncertainty associated with sample variance.
We do not attempt to attribute the amplitude difference
quantitatively to these effects.}

\subsection{Conformity monopoles}
\label{sec:conformity_monopoles}

We next consider the redshift-space conformity monopole \(\Delta f_0(s)\), the
closest observational analogue of the real-space statistic studied in
\citet{PadillaLacernaPaz2026Letter}. Figure~\ref{fig:deltaf0_response} compares
it with the best-fitting \(A_\xi\xi_{\rm mm}^{\rm lin}(s)\) template. The
{central-primary results}, especially DESI, remain close to a
single-amplitude response over most of the fitted range, substantially simpler
than ordinary clustering (Fig. \ref{fig:ordinary_clustering}), {as reflected in the smaller $r_{\min}$
obtained for DESI conformity. SDSS shows a similar trend, although its
larger uncertainties limit the comparison.}

{For conformity, we adopt a fitting range of
$3$--$20\,h^{-1}{\rm Mpc}$ to test the linear template
at smaller separations, where compensation may reduce
nonlinear departures.}

{In the simulations, the $r_{\min}$ values indicate that conformity
does not generally satisfy our quantitative linear-template consistency
criterion down to the smallest measured separations. Nevertheless,}
{under central-primary selection, FLAMINGO, MTNG, and
MDPL2--SAG can show smaller scale-dependent departures
from the linear template than their ordinary clustering
counterparts.} The amplitudes, however, differ
considerably among the galaxy-formation models, much more than for ordinary
clustering. This increased model dependence is expected because the conformity
signal measures differences in clustering between the two colour-selected
populations, and the strength of the relation between colour and clustering is
not the same in the different galaxy-formation models. By contrast, the
difference in amplitude between SDSS and DESI is not markedly larger than for
ordinary clustering. The agreement with a constant linear response is not
perfect and differs among models.
{In the redshift-space simulations (Fig.~3),
all-primary measurements generally show smaller
departures from the fitted linear template at small
separations, while central-primary measurements agree
better at larger separations. Thus, central-primary
selection does not systematically improve agreement
with the linear template.}

\subsection{Projected compensated statistics}
\label{sec:projected_compensated}

We now consider the projected statistics \(\Delta f(r_p)\), \(C_w(r_p)\), and
\(C_{w,\rm auto}(r_p)\). We focus on DESI, whose greater depth yields a larger
effective volume and many more galaxies at fixed number density, allowing a
more precise comparison with the projected and unprojected linear matter
templates. The corresponding SDSS measurements are substantially noisier.

Figure~\ref{fig:projected_compensated} shows that the DESI
{central-primary response ratios} are generally flatter relative to
\(\xi_{\rm mm}^{\rm lin}(r_p)\) than to
\(w_{\rm mm}^{\rm lin}(r_p)\), especially for \(C_w\),
{as indicated by the smaller $r_{\min}$ values for the unprojected
template (dashed arrows).} Thus the compensated residual behaves more like a
local linear response than ordinary projected clustering over the fitted range.

The {lower set of panels} shows the same projected statistics
in the redshift-space simulated catalogues. The simulations do not reproduce
the DESI amplitudes or scale dependences in detail, as expected for
galaxy-formation models not tuned to DESI colour-dependent clustering.
{We therefore restrict the displayed $r_{\min}$ indicators to
the DESI measurements, although the same fitting criterion can be applied
to the simulations.}
However, {central-primary selection} often produces a clearer
\(\xi_{\rm mm}^{\rm lin}\)-like response than {all-primary selection}.
The all-primary ratios again display stronger small-scale deviations,
consistent with their {additional satellite-primary contributions}.

\subsection{Redshift-space quadrupoles and effective projected kernels}
\label{sec:measured_effective_kernels}

{The response ratios above show whether a statistic is well described
by a single-amplitude matter template. The leading-order Kaiser cancellation
and the effective projected kernels discussed in
Sect.~\ref{sec:effective_kernels} provide complementary diagnostics of
the anisotropic and projected responses, respectively.}

{Figure~\ref{fig:quadrupoles} compares the quadrupoles of ordinary
high-colour clustering, $\xi_{HH,2}$, and conformity, $\Delta f_2$,
measured with central-primary selection in DESI and the three simulation
models. In DESI, $\xi_{HH,2}$ shows substantial scale dependence,
changing sign at intermediate separations, whereas $\Delta f_2$
remains close to zero throughout the measured range. MTNG, FLAMINGO,
and MDPL2--SAG reproduce this qualitative behaviour despite differences
in the amplitudes and shapes of their ordinary clustering quadrupoles.}

{The strong suppression of $\Delta f_2$ is consistent with the
leading-order Kaiser cancellation, but extends into separations where
nonlinear clustering and redshift-space effects are important. Thus,
compensation appears to suppress anisotropic contributions beyond the
regime in which the linear derivation strictly applies. This does not
imply the complete removal of redshift-space distortions, since
isotropic contributions to the monopole and higher multipoles may remain.}

{The measured projected conformity statistic is constructed
from ratios of pair counts integrated along the line of sight.
It is therefore not obtained by directly integrating the
unprojected conformity statistic $\Delta f(r_p,\pi)$.
We use the latter only to diagnose how the response is distributed
along the line of sight.}

For projected statistics, the relevant diagnostic is the line-of-sight kernel
\begin{equation}
  \widehat W_S(r_p,\pi)
  =
  \frac{S(r_p,\pi)}
       {A_S\xi_{\rm mm}^{\rm lin}(\sqrt{r_p^2+\pi^2})}.
\end{equation}
Figure~\ref{fig:effective_projection_kernels} shows this
quantity for the {exact, non-linearised unprojected
conformity statistic $\Delta f(r_p,\pi)$}.
{The reference linear matter template has a constant kernel,
$\widehat W_{\rm mm}=1$, whereas ordinary galaxy clustering generally
has a scale-dependent effective response.}
The simulations show that the effective kernel of
$\Delta f(r_p)$ is concentrated preferentially at low line-of-sight separations,
with a coherent positive response extending roughly to
$\pi\sim15$--$20\,h^{-1}{\rm Mpc}$ at small $r_p$. Relative to the
{constant kernel of the linear matter reference template},
this redistribution of weight towards a restricted range of $\pi$
partially counteracts the smoothing introduced by line-of-sight projection.
This provides a geometric explanation for why the projected compensated signal
can resemble $\xi_{\rm mm}^{\rm lin}(r_p)$ more closely than the formally
projected template $w_{\rm mm}^{\rm lin}(r_p)$. DESI shows a weaker positive
response at low $\pi$, while the structure at larger line-of-sight separations
is dominated by noise. The projected cross-compensated statistic $C_w$ has a
qualitatively similar effective kernel and is therefore not shown.

\section{Discussion and conclusions}
\label{sec:conclusions}

{Several compensated combinations of colour-selected
correlations follow single-amplitude linear-matter templates
more closely than ordinary clustering. This is clearest in
DESI BGS, consistent with SDSS MGS, and reproduced
qualitatively by the simulations. Preferential suppression
of nonlinear contributions is a possible explanation,
although the measurements do not identify which modes
are suppressed.}

\begin{figure*}
\centering
\includegraphics[width=0.42\linewidth]{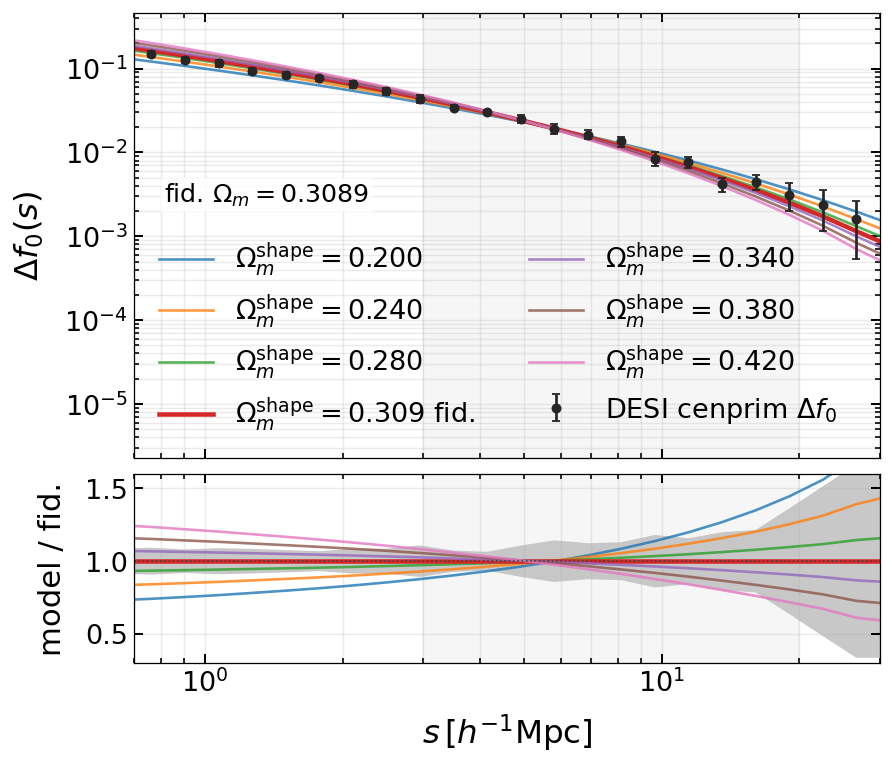}
\includegraphics[width=0.42\linewidth]{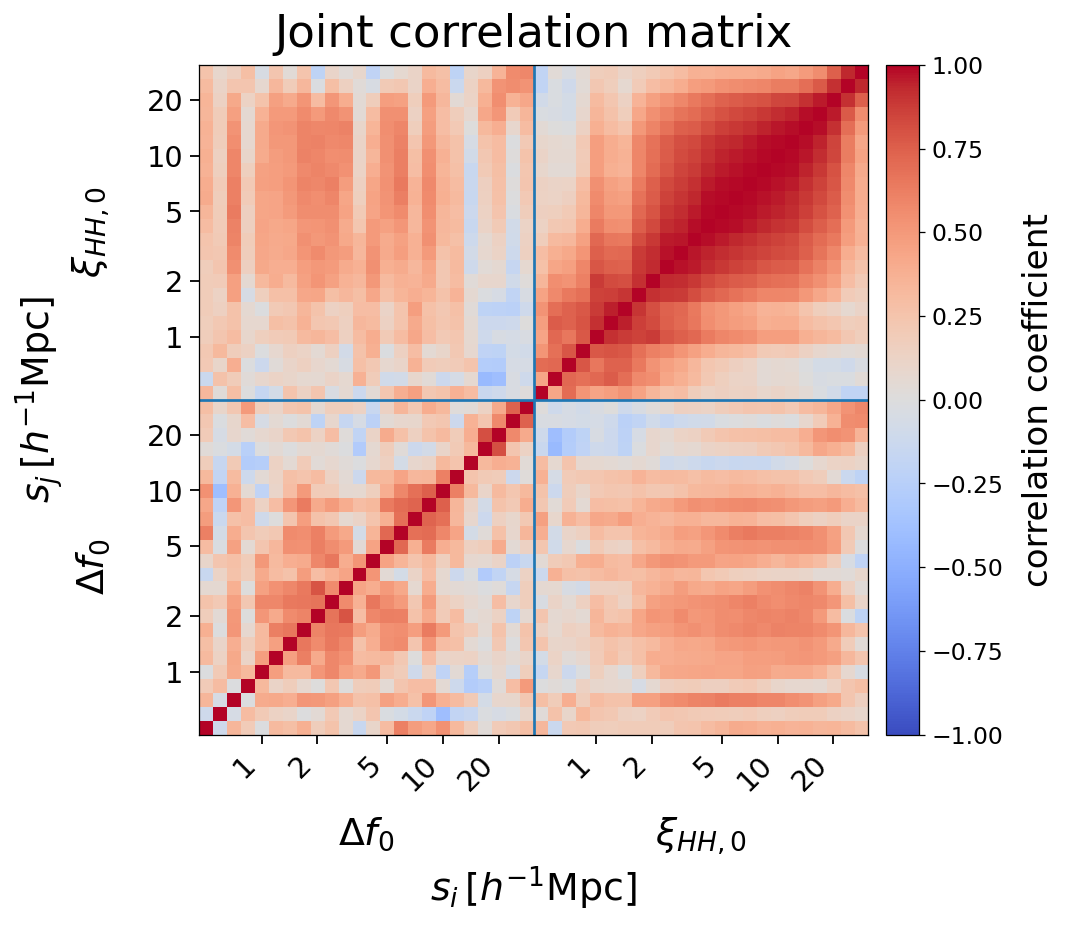}
\caption{
Shape sensitivity and covariance structure of the DESI
{monopole statistics measured with central-primary selection}.
Left: measured conformity monopole $\Delta f_0(s)$
compared with AP-remapped $A_\xi\xi_{\rm mm}^{\rm lin}(s)$ templates obtained
by varying $\Omega_m$ within a fixed flat-$\Lambda$CDM family, with the
amplitude of each template fitted independently. The lower subpanel shows
the templates relative to the fiducial one, with the grey band indicating the
fractional uncertainty of the DESI measurement. Right: joint jackknife
correlation matrix of $\Delta f_0$ and the ordinary high-colour clustering
monopole $\xi_{HH,0}$ for the same DESI
{sample with central-primary selection} at
$n=0.02~h^3{\rm Mpc}^{-3}$. The diagonal blocks show the internal covariance
structure of each statistic, while the off-diagonal blocks show their
cross-correlation. Compared with $\xi_{HH,0}$, $\Delta f_0$ exhibits
substantially weaker long-range correlations between separation bins, while
the non-zero cross-correlation shows that the two statistics retain partially
shared fluctuations.
}
\label{fig:fan_shape_demo}
\end{figure*}

{Two observational limitations remain. Central-primary selection
depends on group-finder classifications, whose uncertainties are not
quantified by our all-primary comparison. In addition, DESI measurements
use a single random catalogue. Although shared random-catalogue
fluctuations are expected to be partially suppressed by compensation,
their residual contribution has not been tested. Both effects require
validation with survey-realistic mocks and additional random catalogues.}

MTNG, FLAMINGO, and MDPL2--SAG produce positive compensated
signals, with model-dependent amplitudes and scale dependence.
{In redshift space, all-primary selection generally
agrees better with the linear template at small separations,
while central-primary selection performs better at larger
separations.} These differences reflect sensitivity to galaxy
occupation and colour assignment, highlighting the need
for survey-calibrated models.

{The fitted amplitude characterises the differential clustering response
of the selected galaxy populations, while departures from the linear
template quantify residual scale dependence. Combined with ordinary
auto- and cross-correlations, these measurements can constrain
assembly-sensitive galaxy occupation without identifying the
conformity signal uniquely with halo assembly bias.}

The shape sensitivity of $\Delta f_0$ is illustrated in the left panel of
Figure~\ref{fig:fan_shape_demo}. The DESI central-primary measurement is
compared with AP-remapped monopole templates obtained by varying $\Omega_m$
within a fixed flat-$\Lambda$CDM family and refitting their amplitudes
independently. The separation among the normalized curves therefore reflects
differences in shape. The AP-remapping procedure and the definition of the
corresponding effective shape parameter are described in
Appendix~\ref{app:omega_shape}. Within this restricted template family,
$\Delta f_0$ retains measurable broad-band shape sensitivity after its
normalization has been marginalized.

{The covariance structure provides complementary evidence that
compensated statistics retain radial information differently from ordinary
clustering.} The right panel of Figure~\ref{fig:fan_shape_demo}
shows the joint jackknife correlation matrix
(${\rm Cov}_{ij}/\sqrt{{\rm Cov}_{ii}{\rm Cov}_{jj}}$)
of $\Delta f_0$ and $\xi_{HH,0}$.
The $\xi_{HH,0}$ block is dominated by broad radial coherence,
whereas the $\Delta f_0$ block is substantially more diagonal.
The cross-blocks nevertheless show coherent stripes: individual $\Delta f_0$
bins remain correlated with fluctuations extending across several
$\xi_{HH,0}$ bins.
{Thus compensation changes the covariance structure without making
the two statistics statistically independent. The more diagonal covariance
suggests a larger effective number of independent radial modes, which we
quantify using the participation-ratio effective rank.} For a correlation matrix $R$ with
eigenvalues $\lambda_i$, we quantify the effective number of independent modes
using the participation-ratio effective rank,
\begin{equation}
r_{\rm eff}(R)=
\frac{\left(\sum_i \lambda_i\right)^2}
{\sum_i \lambda_i^2}.
\label{eq:participation_ratio_rank}
\end{equation}

For the 24-bin monopole measurements, $r_{\rm eff}=6.6$ for $\Delta f_0$ and
$2.3$ for $\xi_{HH,0}$, a factor of $2.9$ gain.
The projected measurements give $7.5$ for $\Delta f(r_p)$ and
$1.7$ for $w_{HH}$, a factor of $4.4$ gain.
{These factors describe the effective rank of the measured
correlation matrices, and not necessarily gains in cosmological information.}

{We estimate the Fisher information on $\Omega_m^{\rm shape}$
from derivatives of the AP-remapped linear templates, using the
joint jackknife covariance of $\Delta f_0$ and $\xi_{HH,0}$
and marginalising over independent amplitudes.
The arrows in Figs.~2 and 3 indicate the minimum scales satisfying
the stable fitting criterion, $11.4\,h^{-1}{\rm Mpc}$
for $\xi_{HH,0}$ and $1.5\,h^{-1}{\rm Mpc}$
for $\Delta f_0$.
For the Fisher calculation, however, we adopt the less
restrictive local criterion ($p\geq0.05$), which gives
$r_{\min}=4.1\,h^{-1}{\rm Mpc}$ for $\xi_{HH,0}$
and $1.5\,h^{-1}{\rm Mpc}$ for $\Delta f_0$.
This allows ordinary clustering to contribute information
from smaller scales, providing a more conservative estimate
of the relative information gain from conformity.
Over their common fitting range, adding conformity
increases the information by approximately $36\%$,
corresponding to a $14\%$ reduction in parameter uncertainty.
Extending $\Delta f_0$ to $1.5\,h^{-1}{\rm Mpc}$,
compared with $4.1\,h^{-1}{\rm Mpc}$ for ordinary clustering,
increases the information by $176\%$
($40\%$ smaller uncertainty).}

{The joint covariance is estimated from 64 jackknife regions and includes the cross-covariance between $\Delta f_0$ and $\xi_{HH,0}$. Its dimension depends on the adopted minimum scales, with at most 48 data bins (24 per statistic). We invert the covariance using a singular value decomposition pseudoinverse with a relative threshold of $10^{-12}$, followed by a Hartlap correction appropriate to the number of bins retained. The resulting Fisher information gains are illustrative.}

{These diagnostics do not establish a gain in cosmological information
beyond standard clustering, BAO, or RSD analyses. In particular,
$\Omega_m^{\rm shape}$ combines transfer-function shape and AP effects
within a restricted template family, while the residual nonlinear
response remains uncalibrated. A joint analysis must establish whether
compensated statistics require fewer nuisance degrees of freedom
than ordinary full-shape clustering.}

{The near-zero conformity quadrupole provides additional evidence
that compensation suppresses the anisotropic response associated with
redshift-space distortions. This is consistent with the cancellation of
the leading Kaiser terms predicted by linear theory. However, the
suppression observed on nonlinear scales is an empirical result and
does not imply that all redshift-space effects have been removed.}
The monopole remains affected by redshift-space distortions, while the
projected statistics retain finite-$\pi_{\rm max}$ and survey-window effects.

The differential construction may also reduce errors common to the compared
galaxy populations, including angular selection fluctuations, radial
incompleteness, fibre-assignment effects, and uncertainty in the overall
clustering normalization. Such cancellation must be established with
survey-realistic mocks. At the same time, the amplitude of the compensated
signal is sensitive to how strongly galaxy properties such as colour modulate
clustering. The differences among the galaxy-formation models seen here
therefore suggest that these statistics can also provide a test of the
connection between galaxy properties and their environments. If confirmed, a
statistic with moderate total signal-to-noise could remain valuable because it
retains more independent radial modes and responds to a different set of
systematic effects.

Our main conclusions are:

\begin{enumerate}

\item
Ordinary colour-selected clustering develops scale-dependent
departures from linear matter templates on small scales,
whereas compensated statistics retain substantially flatter
responses, particularly in DESI with central-primary selection.

\item
{The improved agreement with the linear template
is consistent with preferential suppression of nonlinear
clustering contributions, although we cannot determine
which contributions are suppressed. The effect of primary
selection depends on scale, showing that compensation
does not guarantee linearity.}

\item
{The conformity quadrupole is strongly suppressed relative
to ordinary clustering, consistent with cancellation of the
leading Kaiser terms. This suppression extends empirically
into nonlinear scales but does not eliminate all RSD effects.}

\item
The effective ranks of the compensated covariance matrices
are larger by factors of $2.9$ and $4.4$ for monopole and
projected measurements, respectively. These differences do
not directly imply equivalent gains in cosmological information.

\item
The simulations reproduce the qualitative behaviour but show
substantial differences in amplitude and scale dependence.
Survey-calibrated galaxy models and validated covariances
are required before precision cosmological applications.

\end{enumerate}

The next step is a joint calibration of ordinary and compensated statistics
using empirical Halo Occupation Distribution or Subhalo abundance matching catalogues fitted to DESI colour-dependent
clustering.  Such an analysis should determine whether the residual nonlinear
term in Eq.~\eqref{eq:C_mode_decomp} can be described by a small number
of nuisance degrees of freedom once survey geometry, redshift-space
distortions, and covariance are treated consistently.  A complementary
theoretical direction is to decompose tracer correlations into physically
motivated clustering modes, such as nonlinear mode coupling, halo exclusion,
one-halo structure, and assembly-dependent contributions, and test directly
which mode coefficients are suppressed by the compensated combinations.

\begin{acknowledgements}
NP acknowledges support from PICT Raices Federal 2023-0002. IL acknowledges support from the ANID FONDECYT Regular grant 1261197.
\end{acknowledgements}

\bibliographystyle{aa}
\bibliography{PadillaN}

@article{Norberg2001,
  author  = {Norberg, P. and Baugh, C. M. and Hawkins, E. and Maddox, S. and Peacock, J. A. and Cole, S. and Frenk, C. S. and Bland-Hawthorn, J. and Bridges, T. and Cannon, R. and Colless, M. and Collins, C. and Couch, W. and Dalton, G. and Driver, S. P. and Efstathiou, G. and Ellis, R. S. and Glazebrook, K. and Jackson, C. and Lahav, O. and Lewis, I. and Lumsden, S. and Peterson, B. A. and Sutherland, W. and Taylor, K. and {the 2dFGRS Team}},
  title   = {{The 2dF Galaxy Redshift Survey: luminosity dependence of galaxy clustering}},
  journal = {MNRAS},
  volume  = {328},
  pages   = {64--70},
  year    = {2001},
  doi     = {10.1046/j.1365-8711.2001.04862.x},
  eprint  = {astro-ph/0105500}
}

@article{Norberg2002,
  author  = {Norberg, P. and Baugh, C. M. and Hawkins, E. and Maddox, S. and Madgwick, D. and Lahav, O. and Cole, S. and Frenk, C. S. and Baldry, I. and Bland-Hawthorn, J. and Bridges, T. and Cannon, R. and Colless, M. and Collins, C. and Couch, W. and Dalton, G. and Driver, S. P. and Efstathiou, G. and Ellis, R. S. and Glazebrook, K. and Jackson, C. and Lewis, I. and Lumsden, S. and Peacock, J. A. and Peterson, B. A. and Sutherland, W. and Taylor, K. and {the 2dFGRS Team}},
  title   = {{The 2dF Galaxy Redshift Survey: the dependence of galaxy clustering on luminosity and spectral type}},
  journal = {MNRAS},
  volume  = {332},
  pages   = {827--838},
  year    = {2002},
  doi     = {10.1046/j.1365-8711.2002.05348.x},
  eprint  = {astro-ph/0112043}
}

@article{Hawkins2003,
  author  = {Hawkins, E. and Maddox, S. and Cole, S. and Lahav, O. and Madgwick, D. S. and Norberg, P. and Peacock, J. A. and Baldry, I. K. and Baugh, C. M. and Bland-Hawthorn, J. and Bridges, T. and Cannon, R. and Colless, M. and Collins, C. and Couch, W. and Dalton, G. and De Propris, R. and Driver, S. P. and Efstathiou, G. and Ellis, R. S. and Frenk, C. S. and Glazebrook, K. and Jackson, C. and Jones, B. and Lewis, I. and Lumsden, S. and Percival, W. and Peterson, B. A. and Sutherland, W. and Taylor, K.},
  title   = {{The 2dF Galaxy Redshift Survey: correlation functions, peculiar velocities and the matter density of the Universe}},
  journal = {MNRAS},
  volume  = {346},
  pages   = {78--96},
  year    = {2003},
  doi     = {10.1046/j.1365-2966.2003.07063.x},
  eprint  = {astro-ph/0212375}
}

@article{Zehavi2005,
  author  = {Zehavi, I. and Zheng, Z. and Weinberg, D. H. and Frieman, J. A. and Berlind, A. A. and Blanton, M. R. and Scoccimarro, R. and Sheth, R. K. and Strauss, M. A. and Kayo, I. and Suto, Y. and Fukugita, M. and Nakamura, O. and Bahcall, N. A. and Brinkmann, J. and Gunn, J. E. and Hennessy, G. S. and Ivezi{\'c}, {\v Z}. and Knapp, G. R. and Loveday, J. and Meiksin, A. and Schlegel, D. J. and Schneider, D. P. and Szapudi, I. and Tegmark, M. and Vogeley, M. S. and York, D. G.},
  title   = {{The luminosity and color dependence of the galaxy correlation function}},
  journal = {ApJ},
  volume  = {630},
  pages   = {1--27},
  year    = {2005},
  doi     = {10.1086/431891},
  eprint  = {astro-ph/0408569}
}

@article{Guo2013CMASS,
  author  = {Guo, H. and Zehavi, I. and Zheng, Z. and Weinberg, D. H. and Berlind, A. A. and Blanton, M. R. and Chen, Y. and Eisenstein, D. J. and Ho, S. and Kazin, E. and Manera, M. and Maraston, C. and McBride, C. K. and Nuza, S. E. and Padmanabhan, N. and Parejko, J. K. and Percival, W. J. and Ross, A. J. and Ross, N. P. and Samushia, L. and S{\'a}nchez, A. G. and Schlegel, D. J. and Schneider, D. P. and Skibba, R. A. and Swanson, M. E. C. and Tinker, J. L. and Tojeiro, R. and Wake, D. A. and White, M. and Bahcall, N. A. and others},
  title   = {{The clustering of galaxies in the SDSS-III Baryon Oscillation Spectroscopic Survey: luminosity and color dependence and redshift evolution}},
  journal = {ApJ},
  volume  = {767},
  pages   = {122},
  year    = {2013},
  doi     = {10.1088/0004-637X/767/2/122},
  eprint  = {1212.1211}
}

@article{Peacock2001,
  author  = {Peacock, J. A. and Cole, S. and Norberg, P. and Baugh, C. M. and Bland-Hawthorn, J. and Bridges, T. and Cannon, R. D. and Colless, M. and Collins, C. and Couch, W. and Dalton, G. and Deeley, K. and De Propris, R. and Driver, S. P. and Efstathiou, G. and Ellis, R. S. and Frenk, C. S. and Glazebrook, K. and Jackson, C. and Lahav, O. and Lewis, I. and Lumsden, S. and Maddox, S. and Percival, W. J. and Peterson, B. A. and Price, I. and Sutherland, W. and Taylor, K.},
  title   = {{A measurement of the cosmological mass density from clustering in the 2dF Galaxy Redshift Survey}},
  journal = {Nature},
  volume  = {410},
  pages   = {169--173},
  year    = {2001},
  doi     = {10.1038/35065528},
  eprint  = {astro-ph/0103143}
}

@article{Percival2001,
  author  = {Percival, W. J. and Baugh, C. M. and Bland-Hawthorn, J. and Bridges, T. and Cannon, R. and Cole, S. and Colless, M. and Collins, C. and Couch, W. and Dalton, G. and De Propris, R. and Driver, S. P. and Efstathiou, G. and Ellis, R. S. and Frenk, C. S. and Glazebrook, K. and Jackson, C. and Lahav, O. and Lewis, I. and Lumsden, S. and Maddox, S. and Moody, S. and Norberg, P. and Peacock, J. A. and Peterson, B. A. and Sutherland, W. and Taylor, K.},
  title   = {{The 2dF Galaxy Redshift Survey: the power spectrum and the matter content of the Universe}},
  journal = {MNRAS},
  volume  = {327},
  pages   = {1297--1306},
  year    = {2001},
  doi     = {10.1046/j.1365-8711.2001.04827.x},
  eprint  = {astro-ph/0105252}
}

@article{Tegmark2004PowerSpectrum,
  author  = {Tegmark, M. and Blanton, M. R. and Strauss, M. A. and Hoyle, F. and Schlegel, D. and Scoccimarro, R. and Vogeley, M. S. and Weinberg, D. H. and Zehavi, I. and Berlind, A. and Budavari, T. and Connolly, A. and Eisenstein, D. J. and Finkbeiner, D. and Frieman, J. A. and Gunn, J. E. and Hamilton, A. J. S. and Hui, L. and Jain, B. and Johnston, D. and Kent, S. and Lin, H. and Nakajima, R. and Nichol, R. C. and Ostriker, J. P. and Pope, A. and Scranton, R. and Seljak, U. and Sheth, R. K. and Stebbins, A. and Szalay, A. S. and Szapudi, I. and Verde, L. and Xu, Y. and Annis, J. and Bahcall, N. A. and Brinkmann, J. and Burles, S. and Castander, F. J. and Csabai, I. and Loveday, J. and Doi, M. and Fukugita, M. and Gillespie, B. and Hennessy, G. and Hogg, D. W. and Ivezi{\'c}, {\v Z}. and Knapp, G. R. and Lamb, D. Q. and Lee, B. C. and Lupton, R. H. and McKay, T. A. and Kunszt, P. and Munn, J. A. and O'Connell, L. and Peoples, J. and Pier, J. R. and Richmond, M. and Rockosi, C. and Schneider, D. P. and Stoughton, C. and Tucker, D. L. and Vanden Berk, D. E. and Yanny, B. and York, D. G.},
  title   = {{The three-dimensional power spectrum of galaxies from the Sloan Digital Sky Survey}},
  journal = {ApJ},
  volume  = {606},
  pages   = {702--740},
  year    = {2004},
  doi     = {10.1086/382125},
  eprint  = {astro-ph/0310725}
}

@article{Tegmark2004,
  author  = {Tegmark, M. and Strauss, M. A. and Blanton, M. R. and Abazajian, K. and Dodelson, S. and Sandvik, H. and Wang, X. and Weinberg, D. H. and Zehavi, I. and Bahcall, N. A. and Hoyle, F. and Schlegel, D. and Scoccimarro, R. and Vogeley, M. S. and Berlind, A. and Budavari, T. and Connolly, A. and Eisenstein, D. J. and Finkbeiner, D. and Frieman, J. A. and Gunn, J. E. and Hamilton, A. J. S. and Hui, L. and Jain, B. and Johnston, D. and Kent, S. and Lin, H. and Nakajima, R. and Nichol, R. C. and Ostriker, J. P. and Pope, A. and Scranton, R. and Seljak, U. and Sheth, R. K. and Stebbins, A. and Szalay, A. S. and Szapudi, I. and Verde, L. and Xu, Y. and Annis, J. and Brinkmann, J. and Burles, S. and Castander, F. J. and Csabai, I. and Loveday, J. and Doi, M. and Fukugita, M. and Gillespie, B. and Hennessy, G. and Hogg, D. W. and Ivezi{\'c}, {\v Z}. and Knapp, G. R. and Lamb, D. Q. and Lee, B. C. and Lupton, R. H. and McKay, T. A. and Kunszt, P. and Munn, J. A. and O'Connell, L. and Peoples, J. and Pier, J. R. and Richmond, M. and Rockosi, C. and Schneider, D. P. and Stoughton, C. and Tucker, D. L. and Vanden Berk, D. E. and Yanny, B. and York, D. G.},
  title   = {{Cosmological parameters from SDSS and WMAP}},
  journal = {Phys. Rev. D},
  volume  = {69},
  pages   = {103501},
  year    = {2004},
  doi     = {10.1103/PhysRevD.69.103501},
  eprint  = {astro-ph/0310723}
}

@article{Cole2005,
  author  = {Cole, S. and Percival, W. J. and Peacock, J. A. and Norberg, P. and Baugh, C. M. and Frenk, C. S. and Baldry, I. and Bland-Hawthorn, J. and Bridges, T. and Cannon, R. and Colless, M. and Collins, C. and Couch, W. and Cross, N. J. G. and Dalton, G. and Eke, V. R. and De Propris, R. and Driver, S. P. and Efstathiou, G. and Ellis, R. S. and Glazebrook, K. and Jackson, C. and Jenkins, A. and Lahav, O. and Lewis, I. and Lumsden, S. and Maddox, S. and Madgwick, D. and Peterson, B. A. and Sutherland, W. and Taylor, K.},
  title   = {{The 2dF Galaxy Redshift Survey: power-spectrum analysis of the final data set and cosmological implications}},
  journal = {MNRAS},
  volume  = {362},
  pages   = {505--534},
  year    = {2005},
  doi     = {10.1111/j.1365-2966.2005.09318.x},
  eprint  = {astro-ph/0501174}
}

@article{Sanchez2006,
  author  = {S{\'a}nchez, A. G. and Baugh, C. M. and Percival, W. J. and Peacock, J. A. and Padilla, N. D. and Cole, S. and Frenk, C. S. and Norberg, P.},
  title   = {{Cosmological parameters from cosmic microwave background measurements and the final 2dF Galaxy Redshift Survey power spectrum}},
  journal = {MNRAS},
  volume  = {366},
  pages   = {189--207},
  year    = {2006},
  doi     = {10.1111/j.1365-2966.2005.09833.x},
  eprint  = {astro-ph/0507583}
}

@article{Tegmark2006LRG,
  author  = {Tegmark, M. and Eisenstein, D. J. and Strauss, M. A. and Weinberg, D. H. and Blanton, M. R. and Frieman, J. A. and Fukugita, M. and Gunn, J. E. and Hamilton, A. J. S. and Knapp, G. R. and Nichol, R. C. and Ostriker, J. P. and Padmanabhan, N. and Percival, W. J. and Schlegel, D. J. and Schneider, D. P. and Scoccimarro, R. and Seljak, U. and Seo, H.-J. and Swanson, M. and Szalay, A. S. and Vogeley, M. S. and Yoo, J. and Zehavi, I. and Abazajian, K. and Anderson, S. F. and Annis, J. and Bahcall, N. A. and Bassett, B. and Berlind, A. and Brinkmann, J. and Budavari, T. and Castander, F. and Connolly, A. and Csabai, I. and Doi, M. and Finkbeiner, D. P. and Gillespie, B. and Glazebrook, K. and Hennessy, G. S. and Hogg, D. W. and Ivezi{\'c}, {\v Z}. and Jain, B. and Johnston, D. and Kent, S. and Lamb, D. Q. and Lee, B. C. and Lin, H. and Loveday, J. and Lupton, R. H. and Munn, J. A. and Pan, K. and Park, C. and Peoples, J. and Pier, J. R. and Pope, A. and Richmond, M. and Rockosi, C. M. and Scranton, R. and Sheth, R. K. and Stebbins, A. and Stoughton, C. and Szapudi, I. and Tucker, D. L. and Vanden Berk, D. E. and Yanny, B. and York, D. G.},
  title   = {{Cosmological constraints from the SDSS luminous red galaxies}},
  journal = {Phys. Rev. D},
  volume  = {74},
  pages   = {123507},
  year    = {2006},
  doi     = {10.1103/PhysRevD.74.123507},
  eprint  = {astro-ph/0608632}
}

@article{Baldauf2010,
  author  = {Baldauf, Tobias and Smith, Robert E. and Seljak, Uro{\v s} and Mandelbaum, Rachel},
  title   = {{Algorithm for the direct reconstruction of the dark matter correlation function from weak lensing and galaxy clustering}},
  journal = {Phys. Rev. D},
  volume  = {81},
  pages   = {063531},
  year    = {2010},
  doi     = {10.1103/PhysRevD.81.063531},
  eprint  = {0911.4973}
}

@article{Mandelbaum2013,
  author  = {Mandelbaum, Rachel and Slosar, An{\v z}e and Baldauf, Tobias and Seljak, Uro{\v s} and Hirata, Christopher M. and Nakajima, Reiko and Reyes, Reinabelle and Smith, Robert E.},
  title   = {{Cosmological parameter constraints from galaxy-galaxy lensing and galaxy clustering with the SDSS DR7}},
  journal = {MNRAS},
  volume  = {432},
  pages   = {1544--1575},
  year    = {2013},
  doi     = {10.1093/mnras/stt572},
  eprint  = {1207.1120}
}

@article{Croton2007AssemblyBias,
  author  = {Croton, Darren J. and Gao, Liang and White, Simon D. M.},
  title   = {{Halo assembly bias and its effects on galaxy clustering}},
  journal = {MNRAS},
  volume  = {374},
  pages   = {1303--1309},
  year    = {2007},
  doi     = {10.1111/j.1365-2966.2006.11230.x},
  eprint  = {astro-ph/0605636}
}

@article{Wang2013AssemblyBias,
  author  = {Wang, Lan and Weinmann, Simone M. and {De Lucia}, Gabriella and Yang, Xiaohu},
  title   = {{Detection of galaxy assembly bias}},
  journal = {MNRAS},
  volume  = {433},
  pages   = {515--520},
  year    = {2013},
  doi     = {10.1093/mnras/stt743},
  eprint  = {1305.0350}
}

@article{Zentner2014AssemblyBias,
  author  = {Zentner, Andrew R. and Hearin, Andrew P. and {van den Bosch}, Frank C.},
  title   = {{Galaxy assembly bias: a significant source of systematic error in the galaxy--halo relationship}},
  journal = {MNRAS},
  volume  = {443},
  pages   = {3044--3067},
  year    = {2014},
  doi     = {10.1093/mnras/stu1383},
  eprint  = {1311.1818}
}

@article{Hearin2016DecoratedHOD,
  author  = {Hearin, Andrew P. and Zentner, Andrew R. and {van den Bosch}, Frank C. and Campbell, Duncan and Tollerud, Erik},
  title   = {{Introducing decorated HODs: modelling assembly bias in the galaxy--halo connection}},
  journal = {MNRAS},
  volume  = {460},
  pages   = {2552--2570},
  year    = {2016},
  doi     = {10.1093/mnras/stw840},
  eprint  = {1512.03050}
}

@article{Strauss2002MGS,
  author        = {Strauss, Michael A. and Weinberg, David H. and Lupton, Robert H. and Narayanan, Vijay K. and Annis, James and Bernardi, Mariangela and Blanton, Michael and Burles, Scott and Connolly, Andrew J. and Dalcanton, Julianne and Doi, Mamoru and Eisenstein, Daniel and Frieman, Joshua A. and Fukugita, Masataka and Gunn, James E. and Ivezi{\'c}, {\v Z}eljko and Kent, Stephen and Kim, Rita S. J. and Knapp, Gillian R. and Kron, Richard G. and Munn, Jeffrey A. and Newberg, Heidi Jo and Nichol, Robert C. and Okamura, Sadanori and Quinn, Thomas R. and Richmond, Michael W. and Schlegel, David J. and Shimasaku, Kazuhiro and SubbaRao, Mark and Szalay, Alexander S. and Vanden Berk, Daniel and Vogeley, Michael S. and Yanny, Brian and Yasuda, Naoki and York, Donald G. and Zehavi, Idit},
  title         = {Spectroscopic Target Selection in the Sloan Digital Sky Survey: The Main Galaxy Sample},
  journal       = {The Astronomical Journal},
  year          = {2002},
  volume        = {124},
  number        = {3},
  pages         = {1810--1824},
  doi           = {10.1086/342343},
  eprint        = {astro-ph/0206225},
  archivePrefix = {arXiv}
}

@article{RuizMacias2020BGS,
  author        = {Ruiz-Macias, Omar and Zarrouk, Pauline and Cole, Shaun and Norberg, Peder and Baugh, Carlton and Brooks, David and Dey, Arjun and Duan, Yutong and Eftekharzadeh, Sarah and Eisenstein, Daniel J. and Forero-Romero, Jaime E. and Gazta{\~n}aga, Enrique and Hahn, ChangHoon and Kehoe, Robert and Landriau, Martin and Lang, Dustin and Levi, Michael E. and Lucey, John and Meisner, Aaron M. and Moustakas, John and Myers, Adam D. and Palanque-Delabrouille, Nathalie and Poppett, Claire and Prada, Francisco and Raichoor, Anand and Schlegel, David J. and Schubnell, Michael and Tarl{\'e}, Gregory and Weinberg, David H. and Wilson, Michael J. and Y{\`e}che, Christophe},
  title         = {Preliminary Target Selection for the DESI Bright Galaxy Survey (BGS)},
  journal       = {Research Notes of the American Astronomical Society},
  year          = {2020},
  volume        = {4},
  number        = {10},
  pages         = {187},
  doi           = {10.3847/2515-5172/abc25a},
  eprint        = {2010.11283},
  archivePrefix = {arXiv},
  primaryClass  = {astro-ph.CO}
}

@article{Ross2024DESILSS,
  author        = {Ross, Ashley J. and others},
  title         = {The Construction of Large-scale Structure Catalogs for the Dark Energy Spectroscopic Instrument},
  journal       = {arXiv e-prints},
  year          = {2024},
  pages         = {arXiv:2405.16593},
  eprint        = {2405.16593},
  archivePrefix = {arXiv},
  primaryClass  = {astro-ph.CO}
}

@article{DESI2024SampleDefinitions,
  author        = {{DESI Collaboration} and others},
  title         = {DESI 2024 II: Sample Definitions, Characteristics, and Two-point Clustering Statistics},
  journal       = {arXiv e-prints},
  year          = {2024},
  pages         = {arXiv:2411.12020},
  eprint        = {2411.12020},
  archivePrefix = {arXiv},
  primaryClass  = {astro-ph.CO}
}

@article{BlantonRoweis2007,
  author        = {Blanton, Michael R. and Roweis, Sam},
  title         = {K-Corrections and Filter Transformations in the Ultraviolet, Optical, and Near-Infrared},
  journal       = {The Astronomical Journal},
  year          = {2007},
  volume        = {133},
  number        = {2},
  pages         = {734--754},
  doi           = {10.1086/510127},
  eprint        = {astro-ph/0606170},
  archivePrefix = {arXiv}
}

@article{Cora2018SAG,
  author        = {Cora, Sof{\'\i}a A. and Vega-Mart{\'\i}nez, Cristian A. and Hough, Tom{\'a}s and Ruiz, Andr{\'e}s N. and Orsi, {\'A}lvaro and Mu{\~n}oz Arancibia, Alejandra M. and Gargiulo, Ignacio D. and Collacchioni, Florencia and Padilla, Nelson D. and Gottl{\"o}ber, Stefan and Yepes, Gustavo},
  title         = {Semi-analytic galaxies - I. Synthesis of environmental and star-forming regulation mechanisms},
  journal       = {Monthly Notices of the Royal Astronomical Society},
  year          = {2018},
  volume        = {479},
  number        = {1},
  pages         = {2--24},
  doi           = {10.1093/mnras/sty1131},
  eprint        = {1801.03883},
  archivePrefix = {arXiv},
  primaryClass  = {astro-ph.GA}
}

@article{Pakmor2023MTNGHydro,
  author  = {Pakmor, R. and Springel, V. and Coles, J. P. and Guillet, T. and Pfrommer, C. and Bose, S. and others},
  title   = {The MillenniumTNG Project: The hydrodynamical full physics simulation and a first look at its galaxy clusters},
  journal = {MNRAS},
  volume  = {524},
  pages   = {2539--2555},
  year    = {2023},
  eprint  = {2210.10060},
  archivePrefix = {arXiv},
  primaryClass = {astro-ph.CO}
}

@article{Bose2023MTNGClustering,
  author  = {Bose, S. and Hadzhiyska, B. and Barrera, M. and Springel, V. and White, S. D. M. and others},
  title   = {The MillenniumTNG Project: The large-scale clustering of galaxies},
  journal = {MNRAS},
  volume  = {524},
  pages   = {2579--2594},
  year    = {2023},
  eprint  = {2210.10065},
  archivePrefix = {arXiv},
  primaryClass = {astro-ph.CO}
}

@article{Knebe2018MultiDarkGalaxies,
  author  = {Knebe, A. and Stoppacher, D. and Prada, F. and Behrens, C. and Benson, A. and others},
  title   = {MultiDark-Galaxies: data release and first results},
  journal = {MNRAS},
  volume  = {474},
  pages   = {5206--5231},
  year    = {2018},
  eprint  = {1710.08150},
  archivePrefix = {arXiv},
  primaryClass = {astro-ph.GA}
}

@article{Klypin2016MultiDark,
  author        = {Klypin, Anatoly and Yepes, Gustavo and Gottl{\"o}ber, Stefan and Prada, Francisco and Hess, Steffen},
  title         = {MultiDark simulations: the story of dark matter halo concentrations and density profiles},
  journal       = {Monthly Notices of the Royal Astronomical Society},
  year          = {2016},
  volume        = {457},
  number        = {4},
  pages         = {4340--4359},
  doi           = {10.1093/mnras/stw248},
  eprint        = {1411.4001},
  archivePrefix = {arXiv},
  primaryClass  = {astro-ph.CO}
}

@article{Schaye2023FLAMINGO,
  author = {{Schaye}, Joop and {Kugel}, Roi and {Schaller}, Matthieu and {Helly}, John C. and {Braspenning}, Joey and {Elbers}, Willem and {McCarthy}, Ian G. and {van Daalen}, Marcel P. and {Vandenbroucke}, Bert and {Frenk}, Carlos S. and {Kwan}, Juliana and {Salcido}, Jaime and {Bahé}, Yannick M. and {Borrow}, Josh and {Chaikin}, Evgenii and {Hahn}, Oliver and {Huško}, Filip and {Jenkins}, Adrian and {Lacey}, Cedric G. and {Nobels}, Folkert S. J.},
  title = {{The FLAMINGO project: cosmological hydrodynamical simulations for large-scale structure and galaxy cluster surveys}},
  journal = {Monthly Notices of the Royal Astronomical Society},
  year = {2023},
  volume = {526},
  number = {4},
  pages = {4978--5020},
  doi = {10.1093/mnras/stad2419},
  eprint = {2306.04024},
  archivePrefix = {arXiv},
  primaryClass = {astro-ph.CO}
}

@article{Kugel2023FLAMINGOCalibration,
  author = {{Kugel}, Roi and {Schaye}, Joop and {Schaller}, Matthieu and {Helly}, John C. and {Braspenning}, Joey and {Elbers}, Willem and {Frenk}, Carlos S. and {McCarthy}, Ian G. and {Kwan}, Juliana and {Salcido}, Jaime and {van Daalen}, Marcel P. and {Vandenbroucke}, Bert and {Bahé}, Yannick M. and {Borrow}, Josh and {Chaikin}, Evgenii and {Huško}, Filip and {Jenkins}, Adrian and {Lacey}, Cedric G. and {Nobels}, Folkert S. J. and {Vernon}, Ian},
  title = {{FLAMINGO: calibrating large cosmological hydrodynamical simulations with machine learning}},
  journal = {Monthly Notices of the Royal Astronomical Society},
  year = {2023},
  volume = {526},
  number = {4},
  pages = {6103--6127},
  doi = {10.1093/mnras/stad2540},
  eprint = {2306.05492},
  archivePrefix = {arXiv},
  primaryClass = {astro-ph.CO}
}

@unpublished{PadillaLacernaPaz2026Letter,
  author = {{Padilla}, N. and {Lacerna}, I. and {Paz}, D. },
  title = {{Galactic conformity as a linear response to the matter correlation function}},
  year = {2026},
  note = {Letter in preparation}
}

@article{Planck2013Cosmo,
  author        = {{Planck Collaboration} and Ade, P. A. R. and Aghanim, N. and Armitage-Caplan, C. and Arnaud, M. and Ashdown, M. and Atrio-Barandela, F. and Aumont, J. and Baccigalupi, C. and Banday, A. J. and Barreiro, R. B. and Bartlett, J. G. and others},
  title         = {{Planck 2013 results. XVI. Cosmological parameters}},
  journal       = {Astronomy \& Astrophysics},
  volume        = {571},
  eid           = {A16},
  pages         = {A16},
  year          = {2014},
  doi           = {10.1051/0004-6361/201321591},
  eprint        = {1303.5076},
  archivePrefix = {arXiv},
  primaryClass  = {astro-ph.CO}
}

@article{DESIDR2BAO2025,
  author  = {{DESI Collaboration} and Abdul-Karim, M. and others},
  title   = {DESI DR2 Results II: Measurements of Baryon Acoustic Oscillations and Cosmological Constraints},
  journal = {arXiv e-prints},
  year    = {2025},
  pages   = {arXiv:2503.14738},
  eprint  = {2503.14738},
  archivePrefix = {arXiv},
  primaryClass = {astro-ph.CO}
}

@article{Dressler1980Morphology,
  author        = {Dressler, Alan},
  title         = {Galaxy morphology in rich clusters: implications for the formation and evolution of galaxies},
  journal       = {ApJ},
  volume        = {236},
  pages         = {351--365},
  year          = {1980},
  doi           = {10.1086/157753}
}

@article{Weinmann2006,
  author        = {Weinmann, Simone M. and van den Bosch, Frank C. and Yang, Xiaohu and Mo, H. J.},
  title         = {Properties of galaxy groups in the Sloan Digital Sky Survey: I. The dependence of colour, star formation and morphology on halo mass},
  journal       = {MNRAS},
  volume        = {366},
  pages         = {2--28},
  year          = {2006},
  doi           = {10.1111/j.1365-2966.2005.09865.x},
  eprint        = {astro-ph/0509147},
  archivePrefix = {arXiv},
  primaryClass  = {astro-ph}
}

@article{Wechsler2006,
  author={Wechsler, R. H. et al.},
  title={Dependence of Halo Clustering on Halo Formation History},
  journal={ApJ},
  year={2006},
  volume={652},
  pages={71}
}

@article{GaoWhite2007,
  author={Gao, L. and White, S. D. M.},
  title={Assembly Bias in Dark Matter Halo Clustering},
  journal={MNRAS},
  year={2007},
  volume={377},
  pages={L5}
}

@article{Kauffmann2004Env,
  author        = {Kauffmann, Guinevere and White, Simon D. M. and Heckman, Timothy M. and M\'enard, Brice and Brinchmann, Jarle and Charlot, St\'ephane and Tremonti, Christy and Brinkmann, Jon},
  title         = {The environmental dependence of the relations between stellar mass, structure, star formation and nuclear activity in galaxies},
  journal       = {MNRAS},
  volume        = {353},
  pages         = {713--731},
  year          = {2004},
  doi           = {10.1111/j.1365-2966.2004.08117.x},
  eprint        = {astro-ph/0402030},
  archivePrefix = {arXiv},
  primaryClass  = {astro-ph}
}

@article{sheth2005marked,
  author        = {Sheth, Ravi K. and Connolly, Andrew J. and Skibba, Ramin},
  title         = {Marked correlations in galaxy formation models},
  journal       = {arXiv e-prints},
  year          = {2005},
  eprint        = {astro-ph/0511773},
  archivePrefix = {arXiv},
  primaryClass  = {astro-ph}
}

@article{Lacerna2018Conformity,
  author       = {Lacerna, Ivan and Contreras, Sergio and González, Roberto and Padilla, Nelson and González-Perez, Violeta },
  title        = {Galactic conformity measured in semi-analytic models},
  journal      = {MNRAS},
  year         = {2018},
  volume       = {475},
  number       = {1},
  pages        = {1291--1302},
  doi          = {10.1093/mnras/stx3215},
  eprint       = {1703.10175},
  archivePrefix= {arXiv},
  primaryClass = {astro-ph.GA}
}

@article{skibba2006marked,
  author        = {Skibba, Ramin and Sheth, Ravi K. and Connolly, Andrew J. and Scranton, Ryan},
  title         = {The luminosity-weighted or 'marked' correlation function},
  journal       = {MNRAS},
  volume        = {369},
  number        = {1},
  pages         = {68--76},
  year          = {2006},
  doi           = {10.1111/j.1365-2966.2006.10196.x}
}

@article{Kauffmann2013Conformity,
  author        = {Kauffmann, Guinevere and Li, Cheng and Zhang, Wei and Weinmann, Simone},
  title         = {A re-examination of galactic conformity and a comparison with semi-analytic models of galaxy formation},
  journal       = {MNRAS},
  volume        = {430},
  number        = {2},
  pages         = {1447--1456},
  year          = {2013},
  doi           = {10.1093/mnras/stt007},
  eprint        = {1211.0013},
  archivePrefix = {arXiv},
  primaryClass  = {astro-ph.CO}
}

@article{Hearin2015Beyond,
  author        = {Hearin, Andrew P. and Watson, David F. and van den Bosch, Frank C.},
  title         = {Beyond halo mass: galactic conformity as a smoking gun of galaxy assembly bias},
  journal       = {MNRAS},
  volume        = {452},
  number        = {2},
  pages         = {1958--1969},
  year          = {2015},
  doi           = {10.1093/mnras/stv1358},
  eprint        = {1412.1304},
  archivePrefix = {arXiv},
  primaryClass  = {astro-ph.GA}
}

@article{calderon2018conformity,
  author        = {Calderon, Victor F. and Berlind, Andreas A. and Sinha, Manodeep},
  title         = {Small- and large-scale galactic conformity in SDSS DR7},
  journal       = {MNRAS},
  volume        = {480},
  number        = {2},
  pages         = {2031--2045},
  year          = {2018},
  doi           = {10.1093/mnras/sty2000},
  eprint        = {1712.02797},
  archivePrefix = {arXiv},
  primaryClass  = {astro-ph.GA}
}

@article{pahwa2017conformity,
  author        = {Pahwa, Isha and Paranjape, Aseem},
  title         = {Analytical halo model of galactic conformity},
  journal       = {MNRAS},
  volume        = {470},
  number        = {2},
  pages         = {1298--1313},
  year          = {2017},
  doi           = {10.1093/mnras/stx1325},
  eprint        = {1612.00464},
  archivePrefix = {arXiv},
  primaryClass  = {astro-ph.CO}
}

@article{LacernaPadillaPalma2025,
  author        = {Lacerna, Iv\'an and Padilla, Nelson and Palma, Daniela},
  title         = {Assessing the connection between galactic conformity and assembly-type bias},
  journal       = {A\&A},
  volume        = {703},
  pages         = {A247},
  year          = {2025},
  doi           = {10.1051/0004-6361/202555329},
  eprint        = {2505.03880},
  archivePrefix = {arXiv},
  primaryClass  = {astro-ph.GA}
}

@article{Zehavi2011,
       author = {{Zehavi}, Idit and {Zheng}, Zheng and {Weinberg}, David H. and {Blanton}, Michael R. and {Bahcall}, Neta A. and {Berlind}, Andreas A. and {Brinkmann}, Jon and {Frieman}, Joshua A. and {Gunn}, James E. and {Lupton}, Robert H. and {Nichol}, Robert C. and {Percival}, Will J. and {Schneider}, Donald P. and {Skibba}, Ramin A. and {Strauss}, Michael A. and {Tegmark}, Max and {York}, Donald G.},
        title = "{Galaxy Clustering in the Completed SDSS Redshift Survey: The Dependence on Color and Luminosity}",
      journal = {\apj},
         year = 2011,
        month = jul,
       volume = {736},
       number = {1},
          eid = {59},
        pages = {59},
          doi = {10.1088/0004-637X/736/1/59},
archivePrefix = {arXiv},
       eprint = {1005.2413},
 primaryClass = {astro-ph.CO},
       adsurl = {https://ui.adsabs.harvard.edu/abs/2011ApJ...736...59Z}
}

@ARTICLE{Yang2005GroupFinder,
  author = {{Yang}, Xiaohu and {Mo}, H.~J. and {Jing}, Y.~P. and {van den Bosch}, Frank C.},
  title = "{Galaxy Groups in the 2dFGRS: I. The catalogue and the basic properties}",
  journal = {\mnras},
  year = {2005},
  volume = {356},
  pages = {1293--1307},
  doi = {10.1111/j.1365-2966.2005.08560.x}
}

@ARTICLE{Yang2007SDSSGroups,
  author = {{Yang}, Xiaohu and {Mo}, H.~J. and {van den Bosch}, Frank C. and
            {Pasquali}, Anna and {Li}, Cheng and {Barden}, Marco},
  title = "{Galaxy Groups in the SDSS DR4. I. The Catalogue and Basic Properties}",
  journal = {\apj},
  year = {2007},
  volume = {671},
  pages = {153--170},
  doi = {10.1086/522027},
  eprint = {0707.4640},
  archivePrefix = {arXiv},
  primaryClass = {astro-ph}
}

@ARTICLE{McDonaldSeljak2009,
       author = {{McDonald}, Patrick and {Seljak}, Uro{\v{s}}},
        title = "{How to measure redshift-space distortions without sample variance}",
      journal = {\jcap},
         year = 2009,
        month = oct,
       volume = {2009},
       number = {10},
          eid = {007},
        pages = {007},
          doi = {10.1088/1475-7516/2009/10/007},
archivePrefix = {arXiv},
       eprint = {0810.0323},
 primaryClass = {astro-ph},
       adsurl = {https://ui.adsabs.harvard.edu/abs/2009JCAP...10..007M}
}

@ARTICLE{Hamaus2010,
       author = {{Hamaus}, Nico and {Seljak}, Uro{\v{s}} and {Desjacques}, Vincent},
        title = "{Optimal constraints on local primordial non-Gaussianity from the two-point statistics of large-scale structure}",
      journal = {\prd},
         year = 2011,
        month = mar,
       volume = {84},
       number = {8},
          eid = {083509},
        pages = {083509},
          doi = {10.1103/PhysRevD.84.083509},
archivePrefix = {arXiv},
       eprint = {1104.2321},
 primaryClass = {astro-ph.CO}
}

@ARTICLE{DesjacquesJeongSchmidt2018,
       author = {{Desjacques}, Vincent and {Jeong}, Donghui and {Schmidt}, Fabian},
        title = "{Large-Scale Galaxy Bias}",
      journal = {\physrep},
         year = 2018,
        month = feb,
       volume = {733},
        pages = {1-193},
          doi = {10.1016/j.physrep.2017.12.002},
archivePrefix = {arXiv},
       eprint = {1611.09787},
 primaryClass = {astro-ph.CO},
       adsurl = {https://ui.adsabs.harvard.edu/abs/2018PhR...733....1D}
}

@ARTICLE{MonteroDorta2020,
       author = {{Montero-Dorta}, Antonio D. and {Artale}, M. Celeste and {Abramo}, L. Raul and {Tucci}, Beatriz and {Padilla}, Nelson and {Sato-Polito}, Gabriela and {Lacerna}, Ivan and {Rodriguez}, Facundo and {Angulo}, Raul E.},
        title = "{The manifestation of secondary bias on the galaxy population from IllustrisTNG300}",
      journal = {\mnras},
         year = 2020,
        month = aug,
       volume = {496},
       number = {2},
        pages = {1182-1196},
          doi = {10.1093/mnras/staa1624},
archivePrefix = {arXiv},
       eprint = {2001.01739},
 primaryClass = {astro-ph.GA},
       adsurl = {https://ui.adsabs.harvard.edu/abs/2020MNRAS.496.1182M}
}

\begin{appendix}

\section{Additional compensated combinations}
\label{app:nulls}

This appendix summarizes additional compensated combinations
used as consistency tests. The main text defines the conformity
statistic $\Delta f_A$, the cross-compensated statistic $C_w$,
and the auto-compensated statistic $C_{w,\rm auto}$.
{The latter is identical to $C_w$ under all-primary
selection, as shown in Eq.~\eqref{eq:Cw_auto_def}.}

A further nonlinear consistency statistic is the projected analogue
of the cubic conformity statistic introduced in
\citet{PadillaLacernaPaz2026Letter},
\begin{equation}
G_3^w(r_p)
=
\frac{
\left[w_{Hg}(r_p)-w_{Lg}(r_p)\right]^3
}
{
w_{gg}^2(r_p)
},
\label{eq:app_G3w_def}
\end{equation}
where $g$ denotes the full neighbour sample,
\begin{equation}
w_{Hg}
=
f_A w_{HA}+f_B w_{HB},
\qquad
w_{Lg}
=
f_A w_{LA}+f_B w_{LB},
\label{eq:app_wHg_wLg}
\end{equation}
and $f_A$ and $f_B=1-f_A$ are the neighbour number fractions.
The denominator gives $G_3^w$ the same dimensions as a projected
correlation function and removes two powers of the common parent
clustering amplitude. We use $G_3^w$ only as a consistency test
because it is more nonlinear than $C_w$.

Other compensated combinations can be constructed by subtracting
the parent correlation $w_{gg}$ before taking differences.
For example,
$[w_{Hg}-w_{gg}]-[w_{Lg}-w_{gg}]$
reduces algebraically to $w_{Hg}-w_{Lg}$, and analogous four-term
constructions reduce to $C_w$.
These forms are useful {as algebraic consistency checks
of the cancellation of common terms, although they do not
independently establish cancellation of random-catalogue
or selection-function errors}.
They do not define independent shape statistics.

The correlation of the full parent sample, {including
centrals and satellites}, is
\begin{equation}
w_{gg}
=
f_A^2 w_{AA}
+
2f_Af_B w_{AB}
+
f_B^2 w_{BB},
\label{eq:app_wgg_all}
\end{equation}
where $f_A=f_B=1/2$ for our equal-number colour splits.
{This definition is independent of the primary selection.}

{The auto-compensated statistic can instead be evaluated
using either central-only or all-galaxy colour populations,
with the same selection applied to both members of each pair.
Under central-only selection, $C_{w,\rm auto}$ combines
central--central correlations, whereas under all-galaxy
selection it combines correlations of the full colour-selected
populations. Only in the latter case does
$C_{w,\rm auto}=C_w$.}

The corresponding monopole statistics are obtained by replacing
each projected correlation $w_{XY}(r_p)$ with its redshift-space
monopole $\xi_{XY,0}(s)$. The definitions of $C_0$ and
$C_{0,\rm auto}$ are given in Eqs.~\eqref{eq:C0_def}
and \eqref{eq:C0_auto_def}, respectively.

\section{Effective projected kernels}
\label{app:kernels}

We describe departures from scale-independent linear bias
through the effective response
\begin{equation}
  \xi_{ab}({\bf r})
  =
  b_a b_b\,E_{ab}({\bf r})\,
  \xi_{\rm mm}^{\rm lin}(r),
  \label{eq:Eab_general}
\end{equation}
where $E_{ab}$ collects nonlinear evolution, halo exclusion,
scale-dependent bias, and other departures from the
linear matter response. In real space, $E_{ab}\to1$
on sufficiently large scales. In redshift space, the
linear Kaiser contribution introduces an additional
anisotropic response.

For projected statistics we write
${\bf r}=(r_p,\pi)$ and $s=\sqrt{r_p^2+\pi^2}$.
The cross-compensated correlation is
\begin{equation}
  C(r_p,\pi)
  =
  \xi_{HA}-\xi_{HB}-\xi_{LA}+\xi_{LB},
\end{equation}
where the primary populations $H$ and $L$ are
cross-correlated with the full neighbour populations
$A$ and $B$. Substituting Eq.~\eqref{eq:Eab_general}
gives
\begin{equation}
  C(r_p,\pi)
  =
  {\cal A}_C(r_p,\pi)\,
  \xi_{\rm mm}^{\rm lin}(s),
\end{equation}
with
\begin{equation}
  {\cal A}_C
  =
  b_Hb_AE_{HA}
  -
  b_Hb_BE_{HB}
  -
  b_Lb_AE_{LA}
  +
  b_Lb_BE_{LB}.
  \label{eq:AC_E_rppi}
\end{equation}
The normalized effective kernel introduced in
Sect.~\ref{sec:effective_kernels} is therefore
\begin{equation}
  \widehat W_C(r_p,\pi)
  =
  \frac{{\cal A}_C(r_p,\pi)}{A_C},
\end{equation}
where $A_C$ is the fitted response amplitude.

The exact unprojected conformity field has a related,
but nonlinear, response. For a neighbour property $X$,
\begin{equation}
  \Delta f_X(r_p,\pi)
  =
  \bar f_X
  \left[
  \frac{1+\xi_{HX}}
       {1+\xi_{HN}}
  -
  \frac{1+\xi_{LX}}
       {1+\xi_{LN}}
  \right],
  \label{eq:deltafx_exact_rppi}
\end{equation}
where $N$ denotes the full neighbour sample,
including centrals and satellites. Substitution of
Eq.~\eqref{eq:Eab_general} yields
\begin{equation}
  \Delta f_X(r_p,\pi)
  =
  {\cal A}_X(r_p,\pi)\,
  \xi_{\rm mm}^{\rm lin}(s),
\end{equation}
where
\begin{align}
  \frac{{\cal A}_X}{\bar f_X}
  =
  &
  \frac{
  b_H\left(b_XE_{HX}-b_NE_{HN}\right)
  }
  {
  1+b_Hb_NE_{HN}\xi_{\rm mm}^{\rm lin}(s)
  }
  \nonumber \\
  &
  -
  \frac{
  b_L\left(b_XE_{LX}-b_NE_{LN}\right)
  }
  {
  1+b_Lb_NE_{LN}\xi_{\rm mm}^{\rm lin}(s)
  } .
  \label{eq:AX_E_rppi}
\end{align}

The denominators in this expression account for the
normalisation by the total neighbour density around
each primary population. Consequently, the conformity
response is not simply a linear combination of the
effective responses of the individual correlations.

The ratio construction can suppress contributions
shared by the neighbour populations before the
difference between primary populations is taken.
If the surviving differential response is more
concentrated at small three-dimensional separations
than the common clustering contribution, the
effective kernel of $\Delta f_X$ can be more
localised along the line of sight than that of
ordinary clustering.

We emphasise that the projected conformity statistic
$\Delta f_X(r_p)$ is obtained from ratios of
line-of-sight-integrated pair counts, not by directly
integrating $\Delta f_X(r_p,\pi)$. The unprojected
response and its effective kernel are therefore
diagnostics of the spatial distribution of the
conformity signal, and not an exact projection
formula for the measured $\Delta f_X(r_p)$.

\section{Exploratory effective shape fits}
\label{app:omega_shape}

The template fits used in this paper are not full cosmological likelihoods.
They test whether a statistic follows the shape of the linear matter
correlation function.  If the template shape is varied, the preferred value is
therefore an effective broad-band shape parameter.

For the observed samples, separations are assigned using a fiducial cosmology,
\begin{equation}
r_\perp^{\rm fid}=D_M^{\rm fid}(z)\theta,
\qquad
r_\parallel^{\rm fid}=\frac{c\,\Delta z}{H^{\rm fid}(z)} ,
\end{equation}
where \(D_M(z)\) is the transverse comoving distance.
For a trial cosmology we apply the Alcock--Paczynski remapping
\begin{equation}
r_\perp^{\rm trial}=\alpha_\perp r_\perp^{\rm fid},
\qquad
r_\parallel^{\rm trial}=\alpha_\parallel r_\parallel^{\rm fid},
\end{equation}
with
\begin{equation}
\alpha_\perp(z)=
\frac{D_M^{\rm trial}(z)}{D_M^{\rm fid}(z)},
\qquad
\alpha_\parallel(z)=
\frac{H^{\rm fid}(z)}{H^{\rm trial}(z)} .
\end{equation}
For projected statistics the remapping is applied inside the
\((r_\perp,r_\parallel)\) projection, including the finite
\(\pi_{\rm max}\).  For monopoles we average the isotropic template over
fiducial angle, evaluating it at the corresponding trial-cosmology separation,
\begin{equation}
s_{\rm trial}
=
s_{\rm fid}
\left[
\alpha_\perp^2(1-\mu^2)
+
\alpha_\parallel^2\mu^2
\right]^{1/2}.
\end{equation}

We define \(\Omega_m^{\rm shape}\) as the value of \(\Omega_m\) in a fixed
one-parameter flat-\(\Lambda\)CDM template family that best matches the measured
statistic after this remapping, with the remaining template parameters held
fixed.  This number compresses broad-band transfer-function shape and AP
coordinate remapping into a single effective parameter.  It is not a BAO, AP,
or RSD likelihood.

\begin{figure*}
    \centering
    \includegraphics[width=0.95\linewidth]{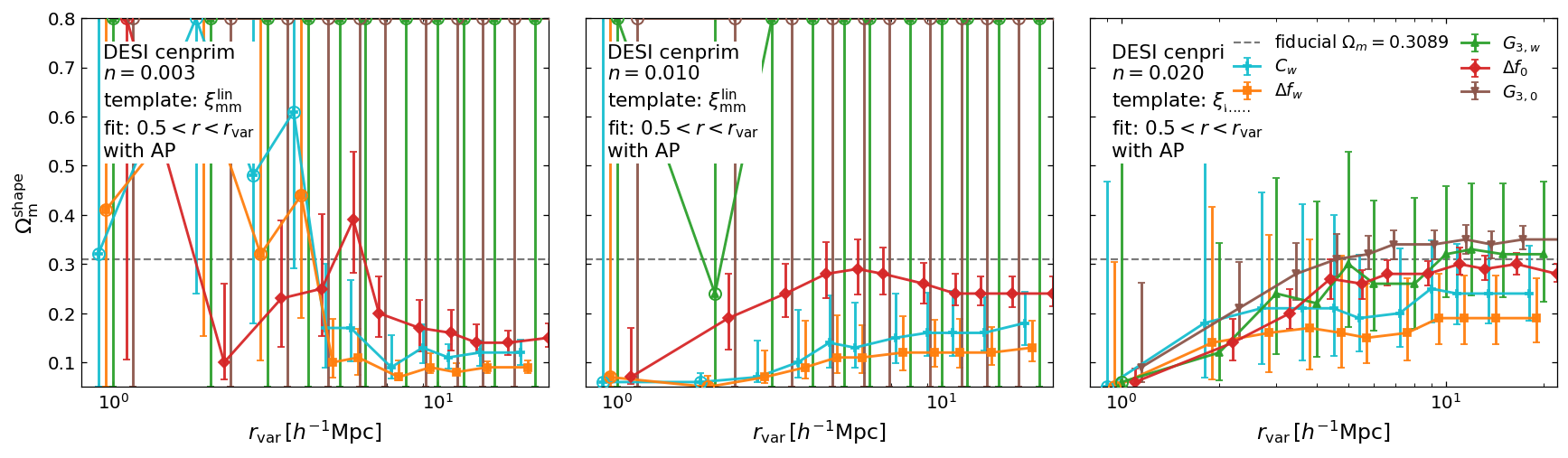}
    \includegraphics[width=0.95\linewidth]{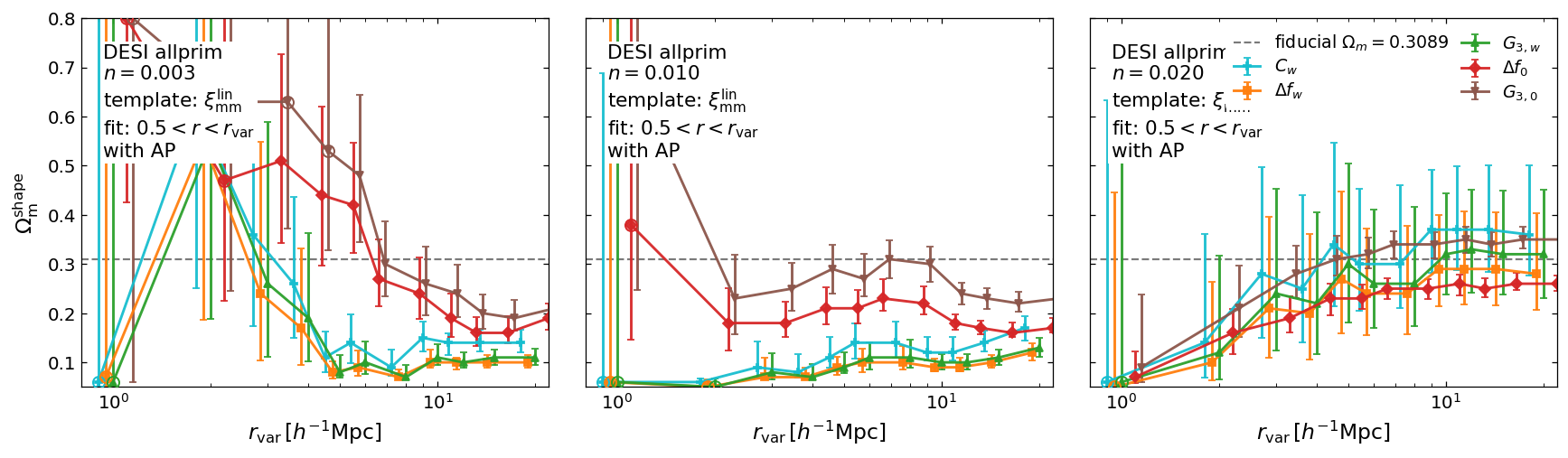}
\caption{
Exploratory DESI \(\Omega_m^{\rm shape}\) fits for the three
fixed-number-density BGS samples. Points show cumulative fits using
AP-remapped \(\xi_{\rm mm}^{\rm lin}\) templates over
\(0.5<r<r_{\rm var}\), where \(r\) denotes \(r_p\) for projected statistics
and \(s\) for monopoles. The dashed horizontal line marks the fiducial
value \(\Omega_m=0.3089\). Upper panels show
{results with central-primary selection}; lower panels show
{results with all-primary selection}.
}
    \label{fig:omshape_n}
\end{figure*}

Figure~\ref{fig:omshape_n} shows exploratory DESI fits for the three
fixed-number-density BGS samples.  The fits use AP-remapped
\(\xi_{\rm mm}^{\rm lin}\) templates and cumulative ranges
\(0.5<r<r_{\rm var}\), where \(r\) denotes \(r_p\) for projected statistics and
\(s\) for monopoles and $r_{\rm var}$ is allowed to vary.  We show both the central-primary and all-primary
definitions because they probe different mixtures of central and satellite
occupation.

The results are not yet stable enough for cosmological interpretation. The
lowest-density sample is dominated by noise and unstable fits. The
intermediate-density sample remains systematically displaced for several
statistics. The densest sample gives the most coherent behaviour, but the
{two primary selections} do not converge to a unique common value.
{Central-primary selection is the default configuration for
conformity, while all-primary selection gives a smoother-looking recovery
for some statistics.}
This difference is itself a warning: the present shape fits remain sensitive to
the primary {selection}, one-halo contributions, and statistic choice.

Fig.~\ref{fig:omshape_n}  shows that compensated conformity statistics contain measurable
broad-band shape information, but also that an honest cosmological analysis
requires survey-tuned mock catalogues, validated covariances, and a model that
jointly describes the ordinary red, blue, and cross-correlations together with
the compensated statistics.

\end{appendix}
\end{document}